\documentclass[preprint,3p, authoryear]{elsarticle}

\usepackage{bbm}
\usepackage{booktabs}
\usepackage{multirow}
\usepackage{float}
\usepackage{tikz}
\usetikzlibrary{arrows.meta,
                decorations.pathreplacing,
                    calligraphy,
                positioning}
\usepackage{mathtools}
\usepackage{amsmath}
\usepackage{bm}
\usepackage{amsfonts,dsfont}
\usepackage{indentfirst}
\usepackage{caption}
\usepackage{subcaption}
\usepackage{graphicx}
\usepackage[table,xcdraw]{xcolor}
\usepackage{todonotes}

\usepackage{url} 
\usepackage{copyrightbox}
\usepackage{hyperref} 
\usepackage{graphicx}
\usepackage{adjustbox}
\usepackage{amsthm}
\usepackage{circuitikz}
\usepackage[T1]{fontenc}
\usepackage{nomencl}
\makenomenclature

\usepackage[
  detect-weight,
  exponent-product=\cdot,
  locale = US,
  group-separator=.,
]{siunitx}
\DeclareSIUnit{\pp}{\textup{p.p.}}

\theoremstyle{definition}

\journal{Energy Economics}

\begin{document}

\begin{frontmatter}

 \author[PWr]{Andrzej Pu{\'c}} \ead{andrzej.puc@pwr.edu.pl}
 \author[PWr]{Joanna Janczura}\ead{joanna.janczura@pwr.edu.pl}
 \affiliation[PWr]{organization={Wroc{\l}aw University of Science and Technology, Faculty of Pure and Applied Mathematics, Hugo Steinhaus Center},
            addressline={Wyb. Wyspia{\'n}skiego 27},
             city={Wroc{\l}aw},
             postcode={50-370},
             country={Poland}}
             

\title{Scenario generation of intraday electricity price paths for optimal trading in continuous markets}

\begin{abstract}
Continuous intraday electricity markets play an increasingly important role in short‑term trading and balancing, yet decision‑making under rapidly evolving price dynamics remains challenging. This paper proposes a comprehensive framework for ensemble forecasting of intraday electricity price trajectories and their translation into adaptive trading decisions. Building on a corrected Support Vector Regression model, the approach extends point predictions to probabilistic trajectory forecasts by introducing scenario generation based on forecast errors of fundamental variables and proposing a novel Support Vector Sorting procedure for the efficient selection of representative scenarios. 
The framework is evaluated using transaction level data from the German intraday continuous market. Empirical results show improvements over benchmark methods in both statistical and economic terms. Fundamental scenarios enhance median trajectory accuracy but produce more concentrated predictive distributions, while historical simulation with scenario selection better captures tail risk. From an economic perspective, ensemble based forecasts outperform na\"{i}ve benchmarks across most of the trading strategies. Dynamic updating through scenario reweighting further improves profitability with limited impact on downside risk. Overall, the results demonstrate that combining kernel based learning with scenario driven uncertainty and adaptive updating provides a flexible and effective approach for forecasting and trading in continuous electricity markets.

\end{abstract}


\begin{highlights}
\item Ensemble forecasts of intraday electricity price trajectories support trading decisions
\item Scenarios based on forecast errors of fundamental variables improve median accuracy
\item Dynamic scenario reweighting enables adaptive intraday trading strategies
\item Ensemble-based forecasts improve trading profitability relative to na\"{i}ve benchmarks
\end{highlights}


\end{frontmatter}


\section{{Introduction}}
The increasing prevalence of renewable energy sources and the growing importance of short-term balancing have substantially enhanced the role of continuous intraday electricity markets. Unlike auction markets, continuous intraday markets operate through a limit order book and allow participants to trade bilaterally on a continuous basis until shortly before delivery. This structure enables market participants to dynamically adjust their positions in response to updated information, including changes in renewable generation, demand conditions, cross-border exchanges, and system imbalances. As a consequence, prices evolve as a sequence of transactions that reflect both fundamental conditions and market microstructure effects. This results in a trading environment characterized by high frequency, rapidly changing liquidity, pronounced volatility, and complex temporal dependencies. 
Although continuous intraday markets have expanded rapidly in recent years -- for example, traded volume at EPEX SPOT increased from 171~TWh in 2024 to 188~TWh in 2025, corresponding to a year‑on‑year growth of approximately 10\% -- the associated forecasting literature remains relatively scarce. In particular, most existing studies focus on auction-based markets or forecasting intraday indices and provide limited insight into the modeling of entire price trajectories in continuous trading environments. This gap is especially relevant for trading applications, where decisions depend not only on expected price levels but also on the temporal evolution and uncertainty of prices throughout the trading horizon.

In this paper, we propose a comprehensive framework for ensemble forecasting of intraday electricity price trajectories and their translation into trading decisions. The approach builds on the corrected Support Vector Regression (cSVR) model proposed recently by \cite{puc_janczura_ijf}, which incorporates the most recent transaction price to adjust its kernel-based representation of the market. {As demonstrated by the authors, this modification improves point forecast accuracy in continuous electricity markets while remaining computationally feasible and allowing for flexible and straightforward adjustments of the model architecture.} We extend this framework along several dimensions. First, the cSVR model is generalized from point to trajectory forecasting and further to ensemble prediction, enabling a probabilistic representation of future price paths. Second, we introduce a scenario construction procedure based on the dynamics of forecast errors of fundamental variables, yielding realistic trajectories that reflect the evolution of market uncertainty over the forecasting horizon. Third, we propose a method for selecting representative scenarios from the full ensemble of predicted paths. Fourth, we develop a model-agnostic approach for dynamically updating trajectory forecasts during the trading horizon. Finally, we conduct an economic evaluation in which the forecasts and their dynamic updates are embedded into trading strategies, allowing us to assess their practical value in terms of profitability and risk.

To implement this framework, we extend the cSVR model from point forecasting to trajectory prediction using two complementary approaches {that were recently applied in the context of day‑ahead electricity price forecasting by \cite{svr_chain_multi_on_dayahead}}: multi‑output regression and regressor chains.  
 Building on these trajectory forecasts, we introduce a scenario generation approach based on the dynamics of forecast errors of the fundamental explanatory variables. {As shown by \cite{uniejewski_ziel_2025}, incorporating distributional information on key fundamentals, such as load and renewable generation, can improve the accuracy of day-ahead price forecasts. In our setting, such information is represented through scenarios of complete trajectories of the forecast errors of fundamental variables}, which naturally reflect the evolution of market uncertainty over the forecast horizon. Importantly, this approach introduces intra-path dependencies in a simple and transparent manner, whereby dependence is captured through the joint evolution of fundamental variables rather than being imposed directly on prices. As a result, the method accounts for the temporal structure of price trajectories while remaining fully data-driven and computationally feasible.  Similar scenario-based approaches have been successfully applied in other forecasting contexts, like macroeconomic indicators \citep{scenarios_macro} or solar power generation \citep{scenarios_solar}.
 We benchmark the proposed approach against ensemble generation based on a classical historical simulation method, in which a sample of residuals is used to construct scenarios, as well as against a standard na\"{i}ve ensemble benchmark used by \cite{NARAJEWSKI2020115801}.

Given the potentially large number of resulting scenarios, the framework is next augmented with a mechanism to improve the efficiency of the ensemble representation. Specifically, we propose a procedure for selecting a representative subset of price‑path scenarios. The method, referred to as Support Vector Sorting (SVS), exploits the parameters of the cSVR model to rank scenarios according to their contribution to the decision function and applies a stopping criterion based on the Wasserstein distance. As a result, the number of scenarios can be substantially reduced while preserving the essential features of the predictive distribution. We show that this reduction in ensemble size can even lead to improvements in the accuracy of probabilistic forecasts. In contrast to alternative scenario selection approaches, like fast forward selection method by \cite{Heitsch2003_fast_forward_selection} or similarity-based procedures \citep[see e.g.,][]{similar_day_load,serafin_nitka_2025_arhnn}, the proposed method is directly linked to the underlying model architecture, which allows for a choice of the most influential scenarios.

The framework is further extended by introducing a novel mechanism for dynamically updating trajectory forecasts during the trading horizon. The scenario weights are adjusted in real time based on the proximity between the realized and predicted price paths observed up to the current trajectory step. The idea is conceptually related to the path shadowing Monte Carlo method of \cite{trajectories_weighting}. We consider two weighting schemes: a kernel-based approach, which provides a flexible adjustment mechanism, and a benchmark method based on an inverse mean absolute error weighting. Both approaches are new in the context of electricity price forecasting, and the model-agnostic nature of the updating procedure makes them applicable even when the underlying forecasting model is not directly accessible. This feature is particularly relevant in practical trading settings, where market participants often rely on externally provided forecasts and lack the ability to update the model structure or generate real‑time forecasts throughout the trading horizon.

Finally, we conduct a case study based on transaction data from the German intraday continuous market to evaluate both the forecasting accuracy and the economic performance of the proposed framework in a trading context. The trajectory forecasts are used to identify optimal transaction timing. We employ decision rules based on the median path forecast as well as band‑based strategies that exploit the ensemble of forecasted price paths. The latter have been applied in a similar trading context by \cite{SERAFIN2022106125,bands_strategy}. The analysis considers two representative types of market participants: sellers and spread (proprietary) traders, with the band‑based strategies additionally examined under both risk‑averse and risk‑seeking preferences. The proposed procedure for dynamically updating price paths allows newly arriving information to be incorporated in real-time into the decision‑making process, leading to adaptive adjustments of optimal transaction times throughout the trading horizon. Moreover, the flexible formulation of the weighting kernel function used for trajectory updates enables the identification of optimal model parameters, facilitating their economic interpretation and providing insight into trading policy design. Strategy performance is evaluated in terms of both profitability and downside risk and is compared with the classical static procedures as well as na\"{i}ve and crystal-ball benchmarks.

The remainder of the paper is organized as follows. Section~\ref{sec:literature} provides an overview of the existing literature on electricity price forecasting in intraday continuous markets. Section~\ref{sec:methodology} introduces the forecasting methodology employed in this study, including the ensemble generation techniques and the scenario selection procedure. The dataset and the forecasting setup are described in Section~\ref{sec:dataset}. Section~\ref{sec:trading_strategies} presents both classical static trading strategies and the proposed dynamically updated approaches. The empirical results are reported and discussed in Section~\ref{sec:case_study}, while Section~\ref{sec:conclusion} concludes the paper.

\subsection{Literature on electricity price forecasting in intraday continuous markets}
\label{sec:literature}
Research on electricity price forecasting in continuous intraday markets remains relatively limited compared to the extensive literature on day-ahead auctions. This gap can be attributed to the later development of continuous trading platforms and to the increased complexity of intraday data, which are irregularly spaced in time and multidimensional due to the evolving order flow. One of the first systematic studies linking auction‑based and continuous markets was provided by \cite{Kath_2018}, who demonstrated that quarter‑hourly day‑ahead prices are strong predictors of the corresponding intraday products. Shortly thereafter, \cite{NARAJEWSKI2020100107} investigated the forecasting of the ID3 index, defined as the volume-weighted average price three hours before delivery, using penalized linear models. While these models did not outperform the na\"{i}ve benchmark for hourly delivery price trajectories, the authors reported  improvements for quarter‑hourly products, which they attributed to lower liquidity relative to the hourly market. Complementary evidence on the role of recent price information was provided by \cite{UNIEJEWSKI2019}, who identified the last observed price as the most influential explanatory variable in the intraday continuous market. Subsequent work by \cite{beating_the_naive} showed that forecasting performance for hourly ID3 products can be enhanced by averaging the most recent observed price with a LASSO‑based forecast. Building on this idea, \cite{puc_janczura_ijf} proposed a SVR approach with kernel correction based on the na\"{i}ve forecast. The method yielded significant improvements in forecasting accuracy for minutely volume‑weighted average prices of quarter‑hourly intraday products for a range of lead times and horizons, outperforming LASSO, standard SVR, and na\"{i}ve benchmark models on the considered EPEX price data.

Beyond point forecasts, the literature has increasingly focused on probabilistic and path‑based forecasting. \cite{NARAJEWSKI2020115801} introduced a framework for simulating intraday price trajectories using generalised additive models, highlighting the role of volatility dynamics and time-to-delivery effects. In parallel, \cite{neural_networks_probabilistic_forecasting} proposed modelling the volume-weighted empirical cumulative distribution function via ensembles of neural networks. The authors showed that it was outperformed by the LASSO model. More recent contributions incorporate richer dependence structures. For instance,  \cite{hirsch2023multivariatesimulationbasedforecastingintraday} model cross-product interactions using copulas and demonstrate that joint modelling across delivery periods improves predictive performance, while \cite{rev2_OMP} show that Bayesian hierarchical approaches with sparse feature selection can outperform both na\"{i}ve and LASSO-based benchmarks. An important complementary line of research focuses on the fundamental drivers of intraday prices and their volatility. \cite{ziel2022elasticity} show that while expected returns are largely unpredictable -- consistent with weak-form market efficiency -- volatility is strongly influenced by the merit-order regime and time to delivery, with tail behaviour driven by trading activity and lagged price differences.

Alongside distributional forecasting, increasing attention has been devoted to representations of price dynamics by generating an ensemble of path forecasts. Methods based on trajectory simulation \citep{NARAJEWSKI2020115801}, Gaussian copulas \citep{SERAFIN2022106125}, and more recently generative machine learning models \citep{bands_strategy}, aim to produce realistic price paths rather than marginal distributions. Such approaches are particularly relevant for decision-making under uncertainty, as they enable the construction of trading strategies based on temporal price patterns. Indeed, \cite{SERAFIN2022106125} demonstrate that trading rules derived from prediction bands based on simulated paths can yield substantial economic gains, while \cite{bands_strategy} extend this paradigm using generative neural networks to improve both forecast realism and trading performance.

Due to the limitations in order book accessibility, most studies rely fully on transaction data. A notable recent exception is \cite{orderfusion_orderbook_encoder}, which proposes an interaction-aware representation of the order book, enabling end-to-end probabilistic forecasting. This direction suggests that integrating market microstructure features may be key to further improving short-term predictive accuracy.

\section{Forecasting methodology}
\label{sec:methodology}
\subsection{Corrected SVR model}
\label{sec:cSVR_path}

Support Vector Regression (SVR) is a kernel-based learning method whose core idea is to approximate an unknown nonlinear relationship by fitting a linear function in a~feature space induced by a kernel. Let $\{(\bm{x}_i, y_i)\}_{i=1}^N$ denote a training sample, where $\bm{x}_i \in \mathbb{R}^d$ is a feature vector and $y_i \in \mathbb{R}$ is the corresponding target value. An intuition behind the SVR is that it seeks a function that is as flat as possible while keeping prediction errors within a prescribed tolerance level. Formally, following \cite{NIPS1996_d3890178}, SVR constructs a linear predictor in a reproducing kernel Hilbert space $\mathcal{H}$,
\begin{equation}
f(\bm{x}) = \bm{w}^\top \phi(\bm{x}) + b,
\end{equation}
where $\phi : \mathbb{R}^d \to \mathcal{H}$ is a feature map associated with a~kernel, $\bm{w} \in \mathcal{H}$ is a weight vector, and $b \in \mathbb{R}$ is an intercept. The model is estimated by minimizing the norm of $\bm{w}$ subject to the requirement that deviations larger than a fixed tolerance $\epsilon$ are penalized linearly. Introducing slack variables $\xi_i, \xi_i^*$ for violations of the $\epsilon$-insensitive tube, the primal optimization problem takes the form \citep{svr_equation}
\begin{equation}
\begin{aligned}
\min_{\bm{w}, b, \bm{\xi}, \bm{\xi}^*} \quad
& \frac{1}{2}\|\bm{w}\|^2
+ C \sum_{i=1}^N (\xi_i + \xi_i^*) \\
\text{subject to} \quad
& \bm{w}^\top \phi(\bm{x}_i) + b - y_i \leq \epsilon + \xi_i, \\
& y_i - \bm{w}^\top \phi(\bm{x}_i) - b \leq \epsilon + \xi_i^*, \\
& \xi_i, \xi_i^* \geq 0, \quad i = 1, \dots, N,
\end{aligned}
\end{equation}
where $C>0$ controls the trade-off between model flatness and tolerance to deviations exceeding $\epsilon$.

Using standard Lagrangian arguments, the problem can be expressed in its dual form. Let $\alpha_i, \alpha_i^*$ denote the Lagrange multipliers associated with the inequality constraints, and let $K(\bm{x}_i,\bm{x}_j)=\langle \phi(\bm{x}_i), \phi(\bm{x}_j)\rangle_{\mathcal{H}}$ denote the kernel function. Following \cite{svr_equation}, the dual optimization problem is given by
\begin{equation}
\begin{aligned}
\min_{\bm{\alpha}, \bm{\alpha}^*} \quad
& \frac{1}{2} (\bm{\alpha} - \bm{\alpha}^*)^\top \bm{Q} (\bm{\alpha} - \bm{\alpha}^*)
+ \epsilon \sum_{i=1}^N (\alpha_i + \alpha_i^*)
+ \sum_{i=1}^N y_i (\alpha_i - \alpha_i^*) \\
\text{subject to} \quad
& \sum_{i=1}^N (\alpha_i - \alpha_i^*) = 0, \\
& 0 \leq \alpha_i, \alpha_i^* \leq C, \quad i = 1, \dots, N,
\end{aligned}
\end{equation}
where $\bm{Q} \in \mathbb{R}^{N \times N}$ is the kernel matrix with entries $\bm{Q}_{ij} = K(\bm{x}_i,\bm{x}_j)$. The resulting SVR predictor can be written as
\begin{equation}
f(\bm{x}) = \sum_{i=1}^N \left(-\alpha_i + \alpha_i^*\right) K(\bm{x}_i, \bm{x}) + b.
\label{eq:cSVR_decision_func}
\end{equation}
Only observations with nonzero values of $\alpha_i$ or $\alpha_i^*$ contribute to the prediction and are referred to as support~vectors.

In this paper, we employ the corrected Support Vector Regression (cSVR) model introduced by \cite{puc_janczura_ijf}. The correction modifies the kernel function by incorporating information from an auxiliary forecast $\hat{y}_i$ associated with each training observation. The corrected kernel is defined as
\begin{equation}
K(\bm{x}_i, \bm{x}_j)
=
\exp\!\left(-l \|\bm{x}_i - \bm{x}_j\|\right)
\exp\!\left(-g \|\hat{y}_i - \hat{y}_j\|^2\right),
\label{eq:csvr_kernel}
\end{equation}
where $l>0$ and $g>0$ are kernel width parameters. The first factor corresponds to a Laplace kernel on the feature space, while the second factor corresponds to a Gaussian kernel applied to the auxiliary forecasts. This construction reduces the influence of training points that are dissimilar either in terms of their features or their associated forecasts, while preserving the computational structure of the standard SVR optimization~problem.

The corrected kernel (\ref{eq:csvr_kernel}) requires specification of width parameters $l$ and $g$. Following \cite{puc_janczura_ijf}, these parameters are determined directly from the empirical distribution of the input data, by matching its empirical $\alpha_2$-quantile to the theoretical $\alpha_1$-quantile  
of the distribution corresponding to the given kernel (Laplace or Gaussian). Specifically, $l = -\frac{\ln(2 - 2\alpha_1)}{\hat{q}_x(\alpha_2)}$ and $g = \frac{z_{\alpha_1}^2}{2 \, [\hat{q}_{\hat{y}}(\alpha_2)]^2}$, where 
$\hat{q}_x(\cdot)$ and $\hat{q}_{\hat{y}}(\cdot)$ denote the empirical quantile functions of the pairwise distances $\{\|\bm{x}_i - \bm{x}_j\|\}_{i,j=1}^N$ and $\{\|\hat{y}_i - \hat{y}_j\|^2\}_{i,j=1}^N$,~respectively.

\subsection{Path forecasting}
\label{sec:path_forecasting_specification}
Let $\bm{y}_t = (y_t(1), y_t(2), \dots, y_t(H)) \in \mathbb{R}^H$ denote a~price path on a forecast horizon of length $H$ and $\hat{\bm{y}_t}$ the corresponding forecast. We extend the cSVR model to this multi‑step setting using two standard approaches: multi‑output regression and the regressor chain method \citep{svr_chain_multi_on_dayahead}.

In the multi-output approach, a~separate SVR model is estimated independently for each forecast horizon,
\begin{equation}
\hat{y}_t(h) = f_h(\bm{x}_t), \qquad h = 1, 2, \dots, H,
\label{eq:multi}
\end{equation}
where each $f_h(\cdot)$ is an SVR model equipped with the corrected kernel (\ref{eq:csvr_kernel}). Note that, in general, the feature vector $\bm{x}_t$ can change with horizon, although in this paper we assume that it is common for the whole path. This approach avoids the accumulation of forecast errors across horizons, but does not explicitly exploit temporal dependencies within the forecast path.

In the regressor chain approach, the forecast path is generated sequentially, following \cite{chain_regression}
\begin{equation}
\begin{aligned}
\hat{y}_t(1) &= f_1\!\left(\bm{x}_t\right), \\
\hat{y}_t(2) &= f_2\!\left(\bm{x}_t, \hat{y}_t(1)\right), \\
\hat{y}_t(h) &= f_h\!\left(\bm{x}_t, \hat{y}_t(1), \ldots, \hat{y}_t(h-1)\right), \qquad h = 3, 4,  \ldots, H.
\end{aligned}
\label{eq:chain}
\end{equation}
Thus, predictions from previous steps, from $\hat{y}_t(1)$ to $\hat{y}_t(h-1)$, are appended to the feature vector and used as additional regressors. This formulation allows the model to capture dependencies along the forecast path at the cost of potential error propagation.

\subsection{Ensemble path forecasting}
\label{sec:prob_path}

Deterministic cSVR models provide path forecasts of future price trajectories. We extend this framework to a probabilistic setting by generating ensembles of future price paths. Two complementary approaches are considered: a~purely data‑driven historical simulation method and a~scenario‑based approach that relies on forecast errors of fundamental variables.

\subsubsection{Historical scenarios}
\label{sec:hist_sim}

The historical simulation framework serves as a natural benchmark, as it requires no additional structural assumptions beyond the estimated cSVR model. Following \cite{alexander2009market}, the method assumes that the empirical distribution of past forecast errors provides a~suitable approximation of the conditional distribution of future errors. The ensemble of forecasted paths is generated by superimposing historical residuals onto the point forecast. Specifically, the $i$-th scenario is defined as:
\begin{equation}
\hat{\bm{y}}^{(i)}_t = \hat{\bm{y}}_t + \bm{e}^{(i)},
\end{equation}
where $i=1, 2, \dots, n$ denotes the scenario index, and $\bm{e}^{(i)}$ is the residual path derived from the training sample:
\begin{equation}
\bm{e}^{(i)} = \bm{y}_i - \hat{\bm{y}}_i.
\end{equation}

\subsubsection{Fundamental scenarios}
\label{sec:fundamental_scenarios}
We propose a fully data‑driven approach to construct an ensemble of path forecasts based on synthetic scenarios of fundamental variables. At each forecast origin $t$, the cSVR model is trained using the information set available at that time. The ensemble of future prices is then generated by repeatedly substituting the unknown future explanatory variable trajectories with its synthetic scenarios derived from the training window. For each scenario index $i$, replacing the unknown future fundamental inputs with a~scenario $\bm{\Delta}^{(i)}_{\mathrm{fs}}$ yields the 
forecast 
\begin{equation} 
\hat{y}^{(i)}_t(h) = f_h\!\left(\bm{x}_t;\bm{\Delta}^{(i)}_{\mathrm{fs}}\right), \label{eq:fs_full_path_forecast}
 \end{equation} 
where the ensemble $\{\hat{y}^{(i)}_t(h)\}_{i=1}^n$ forms the probabilistic forecast for step $h$. Extension to regressor chain is done according to \eqref{eq:chain}. In this way, uncertainty enters the model exclusively through scenario--dependent substitutions of explanatory variables. 

To obtain realistic and data--driven representations of future uncertainty, we construct synthetic fundamental scenarios using only its historical dynamics. 
Instead of relying on the raw realizations of fundamental variables, we characterize uncertainty through differences in forecast errors. This approach isolates the uncertainty actually present at the forecasting time while reducing the influence of seasonality, trends, and temporary or systematic forecast biases. This is particularly relevant for load fundamental variable in our case study; see the first panel of Figure~\ref{fig:why_double_deltas}. 

\begin{figure}[!h]
    \centering
    \includegraphics[width=0.6\textwidth]{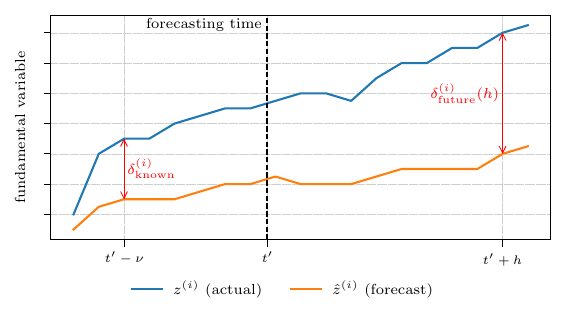}
    \caption{Synthetic example of actual and forecast trajectories for a~fundamental variable in one training window sample. The fundamental scenario from this sample is created by taking the difference between $\delta^{(i)}_{\mathrm{known}}$ and $\delta^{(i)}_{\mathrm{future}}(h)$. Note that the shift between forecasting time and $\delta^{(i)}_{\mathrm{known}}$ is due to data availability delay often present when sourcing the actual values, {see e.g., \cite{entsoe_knowledge_base}}.}
    \label{fig:weather_scenarios_example}
\end{figure}

Let $i$ index the samples in the training window. Denote by $\bm{z}^{(i)}$ the realized values of a fundamental variable and by $\hat{\bm{z}}^{(i)}$ the corresponding forecasts. At forecasting time $t'$, only the historical trajectory of forecast errors is known. We therefore begin by computing the most recent known forecast error
\begin{equation} 
\delta^{(i)}_{\mathrm{known}} = z^{(i)}(t' - \nu) - \hat{z}^{(i)}(t' - \nu), 
\end{equation}
 where $t' - \nu$ is the latest time instant for which both actuals and forecasts are available (due to typical publication delays). Next, for each horizon $h$ 
 within the training window, we compute the future forecast~error
 \begin{equation}
 \delta^{(i)}_{\mathrm{future}}(h) = z^{(i)}(t' + h) - \hat{z}^{(i)}(t' + h).
 \end{equation}
The timeline of these differences is illustrated in Figure \ref{fig:weather_scenarios_example}. Note that future forecast errors can be computed only in the training window.
The synthetic fundamental difference scenario for sample $i$ is then defined as
 \begin{equation}
 \Delta^{(i)}_{\mathrm{fs}}(h) = \delta^{(i)}_{\mathrm{known}} - \delta^{(i)}_{\mathrm{future}}(h). \end{equation} 
Each scenario therefore captures the evolution of forecast errors between the latest available information and future realizations. The full path $\bm{\Delta}_{fs}^{(i)}=\left ( \Delta^{(i)}_{\mathrm{fs}}(1),  \Delta^{(i)}_{\mathrm{fs}}(2),\dots, \Delta^{(i)}_{\mathrm{fs}}(h) \right )$ is subsequently used to generate the corresponding price scenario in~\eqref{eq:fs_full_path_forecast}. This construction naturally preserves intra‑path dependencies at the fundamental‑variable level and satisfies two key features expected of short‑term fundamental scenarios: i) their dispersion increases with the forecast horizon; ii) their distributions remain centered around zero.
Both properties are empirically illustrated for variables used in our case study on the color maps shown in Figure~\ref{fig:scenarios_analysis}.

\subsubsection{Selection of representative scenarios}
\label{sec:scenarios_selection}
Although the impact of seasonality {on fundamental scenarios} is mitigated by considering the forecast errors evolution, these influences cannot be fully eliminated. Moreover, in practical applications, additional factors may affect the relevance of individual {fundamental and} historical scenarios for the current date. For this reason, we argue that sampling uniformly from the entire historical set is suboptimal, as it may introduce unnecessary noise. On the other hand, working with a smaller, well‑selected pool of scenarios enables manual inspection and intuitive validation. This is particularly beneficial in trading contexts, where traders may wish to examine the underlying market situations represented by the scenarios.

We propose a selection method based on the fitted cSVR model structure. Estimating the cSVR model parameters yields dual coefficients $\alpha_i$, $\alpha_i^{\ast}$ for each support vector and for each step of the forecast path. These coefficients appear directly in the cSVR decision function \eqref{eq:cSVR_decision_func}. 
Each historical path contributes support vectors across all steps of the forecast horizon. Treating each path as a~single scenario, we define its importance weight as the sum of the absolute values of the corresponding dual coefficients over the entire forecast path. The scenario weight is then defined as
\begin{equation}
w^{(i)} = \sum_{h=1}^H \left| \left(-\alpha_i^h + {\alpha_i^{\ast}}^h\right) \right|.
\end{equation}
Once the weights $\{w^{(i)}\}_{i=1}^N$ are obtained, historical scenarios are sorted in descending order of $w^{(i)}$. This ranking yields a sequence of scenarios ordered from the most to the least influential for the cSVR model. We refer to this ranking-based selection procedure as Support Vector Sorting (SVS).

To determine how many of the ranked scenarios should be retained, we adopt a stopping rule based on the stabilization of the forecast distribution. This rule ensures that scenarios are kept only if they induce non‑negligible changes in the resulting distribution, yielding a compact yet informative scenario set. Starting from the most influential scenario, additional scenarios are introduced sequentially according to the SVS ranking. 
To quantify the impact of adding the $n$-th trajectory to the current set of $n-1$ scenarios, we compute the first-order Wasserstein distance $\mathcal{W}_1^{(n)}$ \citep[see e.g.,][]{discrete_wasserstein}  between the successive empirical forecast distributions. It captures how much the entire distribution shifts as a consequence of adding the next scenario, providing a natural measure of whether the new scenario meaningfully alters the probabilistic forecast.
We monitor the incremental changes between Wasserstein distances $\Delta^{(n)} = \mathcal{W}_1^{(n-1)} - \mathcal{W}^{(n)}_1$ and compute a moving average of their absolute values over the last 10 iterations. Scenario addition is terminated when this moving average falls below a tolerance threshold $\omega$, i.e.,
$
\frac{1}{10} \sum_{j=n-9}^{n} \left| \Delta^{(j)} \right| < \omega,
$
provided that at least a minimum number of scenarios has already been included. In our implementation, we set $\omega = 0.01$ and require at least 10 scenarios to be selected before stopping is permitted. This safeguard was never binding for the dataset considered. For a~broader discussion on Wasserstein measure use cases and properties see \cite{villani2009_optimal_transport}.

\section{Datasets}
\label{sec:dataset}
\subsection{Electricity price data}

We analyze transaction‑level data from the German continuous intraday electricity market operated by \cite{EPEX}. The market opens after the closure of the day‑ahead and intraday auctions at 15:00~CET and operates continuously until five minutes before delivery. Trading is conducted via a limit order book and covers quarter‑hourly, half-hourly, hourly, and block products. A transaction is executed whenever a buy and a sell order match. The dataset considered in this paper covers quarter-hourly delivery products from 2 January 2019 to 31 December 2020 and includes all transactions for which at least one counterparty was registered in the German market area. Each observation contains a unique trade identifier, transaction timestamp, traded volume, transaction price, delivery period, and market identifier. Self-trades are not~reported.

To obtain a regularly spaced time series suitable for econometric modeling, transaction prices are aggregated into 5-minute intervals using Volume-Weighted Average Prices (VWAPs). This temporal resolution provides a~compromise between preserving high-frequency market dynamics and mitigating microstructure noise, and is commonly adopted in the intraday electricity price forecasting literature, see e.g., \cite{ziel2022elasticity}. For each delivery product, trading starts at 16:00 on the day preceding delivery, which corresponds to the opening time of the continuous intraday market during the period analyzed in this study, and ends five minutes before the start of physical delivery. If no transaction occurs within a given 5-minute interval, the most recent available price is carried forward. Prior to the first transaction for given delivery, the corresponding intraday auction price is used. In the forecasting study, continuous intraday transaction data are used without delay, assuming availability immediately after the end of each 5-minute interval.

\subsection{Forecasting - notation and timeline}
Let us denote the 5-minute VWAP as $P_{d,T}(u)$ and the corresponding volume as $V_{d,T}(u)$, where $u$~is the transaction time in 5-minute intervals, $d$~is the delivery in quarter-hours and $T$~is the day of delivery. 
Our goal is to calculate at a chosen moment $m$ the path price forecast for $u \in \{m+1,\dots,m+31\}$ from 185 minutes to 30 minutes before delivery $d$.

To simplify the forecasting task, we difference the prices with a~lag corresponding to every path step. The resulting differenced price and volume are given by

\begin{equation}
\begin{split}
    &P'_{d, T}(u) = P_{d, T}(u) - P_{d, T}(m),\\&
    V'_{d, T}(u) = V_{d, T}(u) - V_{d, T}(m),
\end{split}
\end{equation}
where $u$ is a 5-minute interval in a~given trajectory. In this way, the differenced variables $P'_{d, T}(u)$ and $V'_{d, T}(u)$ describe the change in price and volume with respect to the forecasting time $m$. 
Finally, we standardize each price trajectory, which is a~standard preprocessing step for fitting the SVR model. Both standardisation and differencing steps are inverted after the forecast calculation. After the inverse transformation, the $i$th scenario path forecast is denoted as $\widehat{P}_{d, T}^{(i)}(u)$.

\subsection{Exogenous variables}
\label{sec:expert_set}

In addition to intraday transaction prices and volumes, we incorporate a rich set of exogenous variables capturing system conditions, cross-border interactions, and market expectations. 
These exogenous variables can be split into four groups:
\begin{itemize}
    \item {Cross-border exchanges}: physical power flows and scheduled commercial exchanges between Germany and its neighbouring markets, i.e. AT, BE, CZ, CH, DK1, DK2, FR, NL, PL, SE4 for physical and AT, CZ, CH, DK1, DK2, FR, NL, PL, SE4 for commercial exchanges. Note that commercial exchanges with BE and NO2 as well as the physical exchange with NO2 are excluded due to missing data in a~significant part of training window. 
    \item {Renewable generation}: actual values and day-ahead forecasts of wind and solar generation in Germany.
    \item {System load}: actual values and day-ahead forecasts of total load in Germany.
    \item {Price information}: day-ahead prices for Germany and neighbouring bidding zones, as well as intraday auction prices for Germany. For the Polish market, prices reported in PLN/MWh between 2017-03-02 and 2019-11-19 are converted to EUR/MWh using historical daily exchange rates sourced from \cite{Investing_EUR_PLN_Historical}.
\end{itemize}

Additionally,  to characterize the local price sensitivity of the intraday market, we employ a merit-order-based elasticity measure derived from the intraday auction order book. The approach follows the methodology proposed by \cite{ziel2022elasticity}, where the auction supply and demand curves are transformed into a representation that allows for a local approximation of the price-volume relationship. Specifically, the anonymous supply and demand curves published after the intraday auction are first transformed into a common curve, called transformed supply curve. This is done by moving the price-dependent bids to the sell side and shifting the volume by the inelastic demand, being the demand at the minimum auction price. As shown by \cite{ziel2022elasticity}, this transformation leaves the equilibrium price unchanged and yields a monotonic supply representation. After constructing the transformed supply curve, the merit-order regime is quantified by evaluating the local slope of this curve around the last known continuous intraday market price for the forecasted delivery. In practice, the slope is approximated using finite differences computed at symmetric volume deviations around the equilibrium point. In line with the empirical analysis by \cite{ziel2022elasticity}, we consider volume perturbations of $\pm 500$, $\pm 1000$, and $\pm 2000$~MW. These magnitudes are chosen to reflect realistic short-term variations in renewable generation. The price bounds used in constructing the transformed curves are restricted to the admissible trading interval of the EPEX SPOT intraday market, namely \([-3000, 3000]\)~EUR/MWh, as specified in the \cite{EPEX_Trading_Brochure_2020} trading documentation.

\begin{figure}[H]
\centering
\begin{tikzpicture}[
    BC/.style = {
        decorate,
        decoration={calligraphic brace, #1,
                    raise=3pt, amplitude=6pt},
                    very thick, thick, 
                    pen colour={black}
                },
    lbl/.style={inner xsep=0pt, text height=1.5ex, text depth=.25ex}    
                        ]
    \tikzstyle{every node}=[font=\tiny]
    \draw[-]   (0,0) -- (0.9,0);

    \node[rectangle, draw=none, minimum size=1pt] at (1.15, 0) {\dots};

    \draw[-]   (1.4,0) -- (4.1,0);

    \node[rectangle, draw=none, minimum size=1pt] at (4.35, 0) {\dots};

    \draw[-]   (4.6,0) -- (6,0);

    \node[rectangle, draw=none, minimum size=1pt] at (6.25, 0) {\dots};

    \draw[-Straight Barb]   (6.5,0) -- (16,0);

    \foreach \x/\i/\j/\k in {
    0.45/X_{d,T}^7/{\scriptstyle\begin{array}{c} \text{10:15} \\ T - 1 \end{array}},
    1.85/X_{d,T}^8/{\scriptstyle\begin{array}{c} \text{13:00} \\ T - 1 \end{array}},
    2.75/X_{d,T}^2/{\scriptstyle\begin{array}{c} \text{14:00} \\ T - 1 \end{array}},
    3.65/X_{d,T}^9/{\scriptstyle\begin{array}{c} \text{15:00} \\ T - 1 \end{array}},
    5.3/{\mathbf{X}_{d,T}^4, X_{d,T}^6}/{\scriptstyle\text{18:00}, T-1}, 11.25/{\mathbf{X}_{T}^1\left(60\left\lfloor \frac{m-181}{60} \right\rfloor \right), \mathbf{X}_{T}^3\left(15\left\lfloor \frac{m - 76}{15} \right\rfloor \right), X_{T}^5\left(15\left\lfloor \frac{m - 76}{15} \right\rfloor \right),
    \mathbf{X}^{10}_{d,T}(m),
    P'_{d,T}(m), V'_{d,T}(m)}/{\scriptstyle m, T}}
    {
        \draw (\x,3pt)    node [lbl, above, yshift=3pt] {$\j$}
            -- ++ 
        (0,-6pt)    node (n\x) [lbl, below] {$\i$};
    }

\end{tikzpicture}
\caption{Timeline of data availability for all exogenous variables used in the study.
}
\label{fig:exog_variab_timeline}
\end{figure}
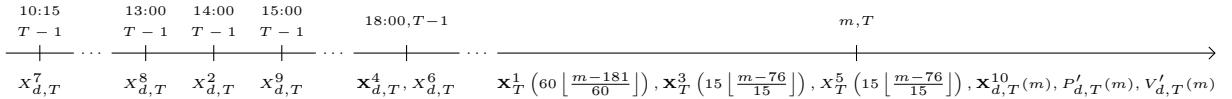

Taking into account data publishing times, the final set of explanatory variables for forecasts calculated at moment $m$ {is presented in Table \ref{tab:variables_table}}. All variables are standardised using the mean and standard deviation computed over the training window to ensure comparability across different scales and units.
\begin{table}[t]
\centering
\begin{adjustbox}{width=0.7\textwidth}
\begin{tabular}{p{10cm} l}
\toprule
Description & Notation \\
\midrule

\multicolumn{2}{l}{\textit{Price variables}} \\
\midrule
Last known intraday price for delivery $d$ & $P_{d,T}(m)$ \\
Vector of intraday price differences for delivery $d$ with lags $1,2,\dots,31$ & $\mathbf{P}'_{d,T}(m)$ \\

\toprule
\multicolumn{2}{l}{\textit{Volume features}} \\
\midrule
Vector of total traded volumes across all deliveries at minute $m$, differenced with lags $1,2,\dots,31$ & $\mathbf{V}'_T(m)$ \\
Sum of traded volumes in the preceding 60 minutes for delivery $d$ & $\sum_{u=m-60}^{m} V_{d,T}(u)$ \\
Number of 5-minute intervals with any transactions in that period & $N_{d,T}(m)$ \\

\toprule
\multicolumn{2}{l}{\textit{Calendar variable}} \\
\midrule
Day of the week variable, a~vector with $D_T=1$ for Monday, ..., $D_T=7$ for Sunday & $D_T$ \\

\toprule
\multicolumn{2}{l}{\textit{Fundamental exogenous variables}} \\
\midrule
Last known actual values of cross-border physical flows & $\mathbf{X}^1_T\left(60\left\lfloor \frac{m-181}{60} \right\rfloor \right)$ \\
Day-ahead scheduled commercial exchanges & $\mathbf{X}^2_{d,T}$ \\
Last known actual wind and solar generations & $\mathbf{X}^3_T\left(15\left\lfloor \frac{m-76}{15} \right\rfloor \right)$ \\
Day-ahead wind and solar generation forecasts & $\mathbf{X}^4_{d,T}$ \\
Last known actual total load & $X^5_T\left(15\left\lfloor \frac{m-76}{15} \right\rfloor \right)$ \\
Day-ahead load forecast & $X^6_{d,T}$ \\
EXAA DE day-ahead price & $X^7_{d,T}$ \\
Day-ahead prices for Germany and neighboring borders & $\mathbf{X}^8_{d,T}$ \\
DE intraday auction price & $X^9_{d,T}$ \\
Vector of merit-order regime measures for three volume deltas & $\mathbf{X}^{10}_{d,T}$ \\

\toprule
\multicolumn{2}{l}{\textit{Fundamental forecast errors}} \\
\midrule
Wind and solar generation forecast errors & $\mathbf{X}^3_T\left(15\left\lfloor \frac{m-76}{15} \right\rfloor \right) - \mathbf{X}^4_{d,T}$ \\
Load forecast error & $X^5_T\left(15\left\lfloor \frac{m-76}{15} \right\rfloor \right) - X^6_{d,T}$ \\
Scheduled commercial exchanges and actual cross-border flows deltas for all borders & $\mathbf{X}^1_T\left(60\left\lfloor \frac{m-181}{60} \right\rfloor \right) - \mathbf{X}^2_{d,T}$ \\

\bottomrule
\end{tabular}
\end{adjustbox}
\caption{A~complete list of explanatory variables used in forecasting study. Vectors are denoted in bold. $d$ and $T$ denote the delivery quarter and day, respectively. Values in brackets represent intraday data availability times. Variables without a time specification are known on the day prior to delivery (see Figure \ref{fig:exog_variab_timeline}). Intraday price and volume data are obtained from \cite{EPEX}, while the remaining data are sourced from \cite{entsoe_transparency_platform}.} 
\label{tab:variables_table}
\end{table}
Note that the hourly granularity cross-border variables use the most recent available value from $60\lfloor (m-181)/60 \rfloor$, while quarter-hourly variables (RES and load) use $15\lfloor (m-76)/15 \rfloor$. The 181 and 76 minutes shifts guarantee that we always use value from the last known full one-hour period. Day-ahead and auction prices, as well as day-ahead RES and load forecasts, are always published before our forecasting times and thus can be used without any delays. Day-ahead prices and forecasts, as well as intraday auction prices, are available before forecasting times and used without delays. The full data availability timeline is shown in Figure~\ref{fig:exog_variab_timeline}.

\section{Trading strategies}
\label{sec:trading_strategies}

Following \cite{SERAFIN2022106125}, we evaluate the economic value of probabilistic electricity price path forecasts in a~fixed-volume scenario. This allows us to assess how the forecasts can be translated into concrete trading decisions for market participants with different objectives. To this end, we propose several trading strategies. All strategies are implemented for two types of agents: a seller, who aims to optimize the timing of a single transaction, and a~spread trader, who seeks to profit from price spreads within the path. Both spread trader and the seller trade 1MWh in each action for every quarter-hourly delivery, which means trading 4MW for each delivery.  
We assume that our trades have a negligible impact on market prices and ignore transaction costs.

We consider two classes of trading strategies. The first is a median‑based strategy, which in its basic form does not account for risk and optimizes solely for profit. The second is a band‑based strategy, which explicitly exploits the dispersion of the predictive distribution and follows the probabilistic band‑trading framework  by \cite{bands_strategy}. Each strategy is constructed in two variants: the static one relies exclusively on the information available at the forecast origin, while the dynamic one allows for intra‑trajectory modifications. Dynamic modifications are introduced by reweighting the forecast paths as new market information becomes available during the delivery~period. 

\subsection{Median-based trading strategy}

The median-based trading strategy relies on the pointwise median of the probabilistic price‑path forecast. Given en ensemble of path scenarios $\widehat{P}^{(i)}_{d,T}(u)$, the median forecast trajectory is obtained as the empirical median computed independently at each step of the path:
\begin{equation}
\widehat{P}^{\mathrm{med}}_{d,T}(u)
=\mathrm{median}\left\{
\widehat{P}^{(i)}_{d,T}(u)
\right\}_{i=1}^n .
\end{equation}
For a~seller, the strategy consists of selecting the path step with the maximum median and executing a~single sell transaction for the delivery period $d, T$ at the corresponding time.
For a spread trader, the median trajectory is used to identify a buy--sell or sell--buy sequence. Specifically, the trader determines the minimum and maximum of \(\widehat{P}^{\mathrm{med}}_{d,T}(u)\) over the forecast path and executes the corresponding opening and closing transactions depending on their temporal ordering. This yields a deterministic trading plan constructed solely from the median forecast trajectory.

\subsection{Band-based trading strategy}
Following \cite{bands_strategy}, we apply a trading strategy based on prediction bands that account for temporal dependence along the forecast path by enforcing a Simultaneous Coverage Probability (SCP). The upper prediction band $\bm{B}^{U}_{d,T}$ is defined so that the entire $i$th price trajectory scenario lies below the band with probability equal to the SCP
\begin{equation}
\mathbb{P}\!\left(
\widehat{P}^{(i)}_{d,T}(u) \le B^{U}_{d,T}(u), \; \forall u
\right)
=
\mathrm{SCP}.
\label{eq:upper_band}
\end{equation}
Analogously, the lower prediction band $\bm{B}^{L}_{d,T}$ satisfies
\begin{equation}
\mathbb{P}\!\left(
B^{L}_{d,T}(u) \le \widehat{P}^{(i)}_{d,T}(u), \; \forall u
\right)
=
\mathrm{SCP}.
\label{eq:lower_band}
\end{equation}
The prediction bands are constructed using the filtering procedure proposed by \cite{STASZEWSKA2007121}. For the upper band, simulated price paths are removed on the basis of their pathwise maxima until only a fraction corresponding to the desired SCP remains. The upper band is then obtained as the pointwise maximum across the remaining trajectories. For the lower band, paths are filtered according to their pathwise minima, and the band is constructed as the pointwise minimum of the retained trajectories. This procedure ensures that the resulting bands satisfy the simultaneous coverage property and preserves the temporal dependence structure of the simulated paths.

The upper prediction band represents the highest price that can be expected with probability SCP, while the lower band represents the lowest such price. For a risk-seeking seller strategy, the selling time is chosen as the subperiod at which the upper prediction band reaches its maximum value. For a risk-averse seller strategy, the selling time is chosen as the subperiod at which the lower prediction band reaches its maximum value, thus maximizing the worst-case price consistent with the prescribed SCP. Analogously, the risk-seeking spread trader will buy at minimum of lower band and sell at maximum of upper band, while the risk-averse will buy at maximum of lower band and sell at minimum of upper band. To our best knowledge, this is the first attempt~of applying a band-based strategy to spread trading in the electricity forecasting~literature.

\subsection{Intra-trajectory {forecast update}}
\label{subsec:intra_traj_median}

We assume that the initial trading plan, based on the cSVR forecast, can be modified in later steps of the trajectory, as new market information arrives. Instead of repeating forecasts calculations, we utilize this new information in a manner similar to \cite{trajectories_weighting}. 
At each step, we compare the observed price path with the individual scenario trajectories and assign weights according to their proximity to the realized values. The weight for the $i$-th scenario at time $\tau$ is calculated using a generalized Gaussian kernel:
\begin{equation}
  K^{(i)}_{d,T,\tau} =
  \exp\!\left(
    -\,MAE_{d,T,\tau}\,
    \left(
      \sum_{u=1}^{\tau} w_{\tau,u}\,\left| P_{d,T}(u) - \widehat{P}^{(i)}_{d,T}(u) \right|^{2}
    \right)^{\!p/2}
  \right),
  \label{eq:kernel_weight}
\end{equation}
where $p$ is the shape parameter of the kernel and $w_{\tau,u}$ are exponentially decaying time weights
\begin{equation}
  w_{\tau,u} =
  \frac{\exp\!\bigl(-\lambda(\tau - u)\bigr)}
       {\displaystyle\sum_{s=1}^{\tau}\exp\!\bigl(-\lambda(\tau - s)\bigr)},
\end{equation}
with $\lambda$ controlling the decay rate. We impose stricter penalization when the realized path deviates markedly from the initially forecasted median trajectory by defining the kernel scale as
\begin{equation}
  MAE_{d,T,\tau}
  = \frac{1}{\tau}\sum_{u=1}^{\tau}
    \left| P_{d,T}(u) - \widehat{P}^{\mathrm{med}}_{d,T}(u) \right|.
\end{equation}
In order to prevent a degenerate case $MAE_{d,T,\tau}=0$, which can occur early in the path, we apply the lower bound of $0.01$ EUR/MWh. 

\begin{figure}[!h]
    \centering
    \includegraphics[width=\textwidth]{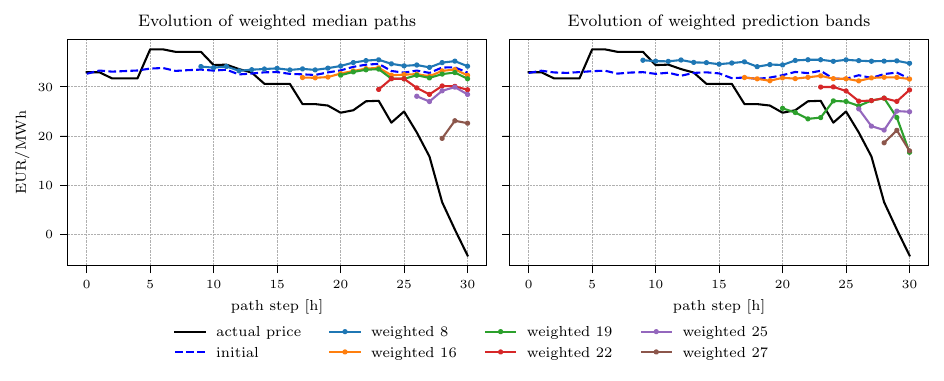}
    \caption{Example of the evolution of weighted median paths and bands under $\mathbb{P}_K$. {The initial median path {forecast} and {prediction} band are marked with dashed blue lines.} Both median and band use the same parameters for {weighting} kernel, i.e. $p=0.5$ and $\lambda=0.35$, resulting from grid search performed in case study. The band is upper with SCP$=5\%$.}
    \label{fig:weighted_paths_example}
\end{figure}

The weights $K^{(i)}_{d,T,\tau}$ assigned to each scenario are used to update the path forecasts. Specifically, both the median and the prediction bands are re-calculated at every step of the trajectory using the updated scenario distribution, in which scenarios are no longer equally weighted but instead carry probabilities proportional to their respective weights. Before being applied, all weights computed for the scenario paths are normalized to sum to one. Let $\mathbb{P}_K$ denote the probability measure induced by the normalized weights $K^{(i)}_{d,T,\tau}$. The updated median path forecast, $\widehat{P}^{\mathrm{med_w}}_{d,T,\tau}(u)$, is then obtained as the weighted median of the remaining future parts of the simulated trajectories under $\mathbb{P}_K$. Similarly, the adjusted weighted upper and lower prediction bands, $B^{U,\mathrm{w}}_{d,T,\tau}(u)$, $B^{L,\mathrm{w}}_{d,T,\tau}(u)$ for future times $u>\tau$  are defined as:
\begin{equation}
\mathbb{P}_K\!\left(
\widehat{P}^{(i)}_{d,T}(u) \le B^{U,\mathrm{w}}_{d,T,\tau}(u), \; \forall\, u > \tau
\right)
=
\mathrm{SCP},
\end{equation}
\begin{equation}
\mathbb{P}_K\!\left(
B^{L,\mathrm{w}}_{d,T,\tau}(u) \le \widehat{P}^{(i)}_{d,T}(u), \; \forall\, u > \tau
\right)
=
\mathrm{SCP}.
\end{equation}
 An illustration of the evolution of the weighted median and the corresponding bands along the trajectory is shown in Figure~\ref{fig:weighted_paths_example}.

As a benchmark to kernel‑based weighting, we also consider a scheme that weights scenarios according to their MAE up to the current step. This approach follows the logic commonly used in the forecast combination literature, introduced by \cite{averaging_forecasts}. Specifically, the weight assigned to the $i$th scenario at time $\tau$ is inversely proportional to its average absolute deviation from the observed path
\begin{equation}
w^{(i)}_{d,T,\tau} =
\frac{\left(\frac{1}{\tau}\sum_{u=1}^{\tau}
\left|P_{d,T}(u)-\widehat{P}^{(i)}_{d,T}(u)\right|\right)^{-1}}
{\displaystyle\sum_{j=1}^{n}
\left(\frac{1}{\tau}\sum_{u=1}^{\tau}
\left|P_{d,T}(u)-\widehat{P}^{(j)}_{d,T}(u)\right|\right)^{-1}}.
\end{equation}
Thus, trajectories that follow the realized price path more closely receive larger weights. In contrast to the kernel‑based approach, this benchmark requires no additional tuning parameters and therefore serves as a simple, nonparametric reference for the dynamic scenario reweighting procedure.

\subsection{Dynamically adapted strategies}
\label{subsec:intra_traj_median_strat}

The updated forecast is dynamically incorporated into the trading strategy. For clarity, we present the procedure for the buy–sell spread trader strategy based on the median forecast, as the other cases, i.e. sell-buy and seller strategies, are analogous. Let the current step be $\tau$ and assume that the current plan is to enter a long position (buy) at $u_0 > \tau$ and close it (sell) at $u_1 > \tau$. Using the updated median path $\widehat{P}^{\mathrm{med_w}}_{d,T,\tau}(u)$ we assess whether modifying the planned entry and exit times increases the expected return. The adjustment is implemented only if the gain implied by the revised entry $u_0^*$ and exit $u_1^*$ exceeds that of the initial plan by more than a~chosen threshold $\eta$:
\begin{equation}
    \widehat{P}^{\mathrm{med_w}}_{d,T,\tau}(u_1^{*}) - \widehat{P}^{\mathrm{med_w}}_{d,T,\tau}(u_0^{*}) - \eta >  \widehat{P}^{\mathrm{med_w}}_{d,T,\tau}(u_1) - \widehat{P}^{\mathrm{med_w}}_{d,T,\tau}(u_0).
\label{eq:updated_entry_exit}
\end{equation}
When $u_0=\tau$, the position is opened immediately, so the realized price $P_{d,T}(u_0)$ replaces the forecast {in \eqref{eq:updated_entry_exit}}.

After entering position at $u_0$, the strategy remains adaptive. In this case, the position is closed earlier at $u_1^* < u_1$ if it yields a higher profit than the profit forecasted under the initial trading plan:
\begin{equation}
     P_{d,T}(u_1^*) - P_{d,T}(u_0) - \eta > \widehat{P}^{\mathrm{med_w}}_{d,T,\tau}(u_1) - P_{d,T}(u_0).
    \label{eq:take_profit_earlier}
\end{equation}
Analogously, at the scheduled exit time $u_1$ we postpone closing the position to $u_1^* > u_1$ if the updated median forecast indicates a higher future profit
\begin{equation}
    \widehat{P}^{\mathrm{med_w}}_{d,T,\tau}(u_1^*) - P_{d,T}(u_0) - \eta > P_{d,T}(u_1) - P_{d,T}(u_0),
    \label{eq:take_profit_later}
\end{equation}
where $u_1^{*}$ corresponds to the maximum value from the updated median trajectory forecast. If the strategy is rescheduled at time $\tau$, the updated times $u_0^*$ and $u_1^*$ are adopted as the new plan and used in subsequent iterations.

The threshold $\eta$ ensures that only meaningful improvements in expected profitability trigger deviations from the initial trading plan. 
{We consider four alternative specifications of $\eta$, all based on the residuals of the {initially} forecasted price path observed up to the current update step. From the most to the least extreme we consider 3 standard deviations of the residual vector ($3\sigma$), the 5--95 interpercentile range (5--95 IPR), the interquartile range (IQR) and the mean absolute error (MAE). These measures differ in the extent to which they penalize large deviations: $3\sigma$ is highly sensitive to extreme outliers, whereas the IPR, IQR, and MAE increasingly reflect the typical magnitude of forecasting errors rather than rare shocks. Making $\eta$ dependent on the empirical distribution of residuals allows the strategy to adjust the required improvement threshold to the prevailing forecast uncertainty -- higher residual variability leads to a larger $\eta$, preventing overreaction to noise, while lower variability reduces $\eta$, enabling the strategy to respond more readily when the forecast becomes more precise. Note that if only a~single residual observation is available in the path so far, each threshold is replaced by its absolute value.} 

Although the strategy is formulated using realized prices for clarity, it is naturally adaptable to continuous markets where execution depends on limit order placement. In practice, the forecasted median trajectory could guide the optimal timing and price levels of orders by comparing immediate execution gains against potential future returns. The specific limit order levels would in such context reflect the trader’s incentives to execute transaction within the current time step.

\section{Case study}
\label{sec:case_study}
\subsection{Ensemble path forecasts}
We employ the cSVR model as defined in Section \ref{sec:cSVR_path}.
The parameters of the kernel widths $l$ and $g$ in formula \eqref{eq:csvr_kernel} are set based on heuristics discussed broader by \cite{puc_janczura_ijf}. For the Laplace kernel, we use $\alpha_1 = 0.75$ and $\alpha_2 = 0.5$, while 
for the Gaussian kernel, a wider shape is preferred to avoid overly strong local correction; therefore, we use $\alpha_1 = 0.75$ and $\alpha_2 = 0.75$. 
In addition to kernel specification, the cSVR requires two hyperparameters: the regularisation parameter $C$ and the insensitive loss parameter $\epsilon$. In the empirical study, we use the \texttt{scikit-learn} implementation \citep[][ version 1.3.0]{scikit-learn} with $C = 1$ and $\epsilon = 0.1$, corresponding to the default settings. Systematic cross-validation over these parameters would entail a~substantial computational burden given the scale of the forecasting task. Preliminary experiments indicated that the forecast improvements from hyperparameter tuning are marginal relative to the associated increase in computational time. Using cSVR model defined in such a~way, we apply and compare both multi-output \eqref{eq:multi} and chain regressor \eqref{eq:chain} forecasting frameworks.

We employ both historical simulation and fundamental scenarios for probabilistic paths generation. Fundamental scenarios are defined as in Section \ref{sec:fundamental_scenarios} and calculated for system load, solar and wind generation fundamental variables based on the forecasts published by the transmission system operator. Their values are evaluated at 15-minute intervals along the forecast horizon. This choice is dictated by variables granularity which is greater than this of the preprocessed prices.  {An illustration of the fundamental scenarios distribution is plotted in Figure \ref{fig:scenarios_analysis}}. The set of scenarios is additionally augmented by splitting positive and negative deviations into separate explanatory variables, effectively doubling the number of scenario dimensions and allowing the model to capture asymmetric effects. In the historical simulation method the in-sample residues paths are used for scenarios generation.

As a~na\"{i}ve benchmark for the cSVR model, we also employ an empirical trajectory resampling procedure following \cite{NARAJEWSKI2020115801}. Historical intraday increment trajectories observed in the training window are treated as empirical scenarios. A~fixed number of 1000 such scenarios is drawn with replacement from the training window and cumulatively added to the last observed price at the current forecast origin. This approach preserves the empirical cross-horizon dependence structure of price changes without imposing parametric assumptions and provides a~natural benchmark for evaluating model-based probabilistic trajectory forecasts.

Forecasts are generated in an expanding window scheme. The initial training window spans from 2 January 2019 to 2 July 2019. The models estimated in this window are used to generate forecasts for 3 July 2019. The window is then expanded by one day and the procedure is repeated sequentially. 
The test window for both the path forecasts accuracy and the evaluation of trading strategies spans from January 1 to December 31, 2020. 
Finally, we also compare three windows for scenarios choice for the na\"{i}ve and both historical and fundamental cSVR scenarios. First window spans over the whole training window, so it grows over time with its expansion. The next two windows are limited to 28 and 182 days. The goal of this comparison is to assess the impact of artificially choosing samples closer to the current market regime.

The median trajectory forecast accuracy is evaluated using the MAE (Mean Absolute Error) while the CRPS (Continuously Ranked Probability Score) is used for the probabilistic evaluation of the generated ensemble. Formally, the MAE is calculated as

\begin{equation}
    MAE = \frac{1}{96 \cdot 31\cdot N_{\mathrm{eval}}} \sum_{d=1}^{96} \sum_{u=1}^{31} \sum_{T=1}^{N_{\mathrm{eval}}} \left | P_{d,T}(u) -\widehat{P}^{\mathrm{med}}_{d,T}(u) \right |,
    \label{eq:MAE}
\end{equation}
where $96$ is the umber of deliveries, $31$ is the number of path steps, and $N_{\mathrm{eval}}$ is the length of test window. The CRPS, following, e.g., \cite{Gneiting01032007} and \cite{NOWOTARSKI20181548}, is approximated by the Pinball-Score (PB) evaluated on a dense equidistant grid of quantile levels $\mathcal{A} = \{0.01,\dots,0.99\}$, i.e.
\begin{equation}
\mathrm{CRPS}_{d,T}(u)
=
\frac{1}{|\mathcal{A}|}
\sum_{\alpha \in \mathcal{A}}
\mathrm{PB}_{d,T}^{\alpha}(u),
\end{equation}
 where $\mathrm{PB}_{d,T}^{\alpha}(u)$ denotes the pinball loss at quantile level $\alpha$ defined as
\begin{equation}
\mathrm{PB}^{\alpha}_{d,T}(u)
=
\begin{cases}
(1-\alpha)\left(\hat{q}_{d,T}^{\alpha}(u) - P_{d,T}(u)\right), & \text{if } P_{d,T}(u) < \hat{q}_{d,T}^{\alpha}(u), \\
\alpha \left(P_{d,T}(u) - \hat{q}_{d,T}^{\alpha}(u)\right), & \text{otherwise}
\end{cases}
\end{equation}
with $\hat{q}_{d,T}^{\alpha}(u)$ being the empirical $\alpha$-quantile of the predictions ensemble. The overall CRPS is obtained by averaging over all days in the evaluation window and all path steps:
\begin{equation}
\mathrm{CRPS}
=
\frac{1}{96 \cdot 31\cdot N_{\mathrm{eval}}} \sum_{d=1}^{96} \sum_{u=1}^{31} \sum_{T=1}^{N_{\mathrm{eval}}}
\mathrm{CRPS}_{d,T}(u).
\label{eq:CRPS}
\end{equation}

\subsubsection{Results}

The obtained values of MAE and CRPS of each model configuration are reported in Table \ref{tab:CRPS_MAE_results}. We first examine the effect of restricting the scenario sampling window to 28 and 182 days. The results indicate that such restrictions yield improvements only in configurations where SVS is applied, and only in terms of MAE. However, a~substantial reduction of the window to 28 days leads to a~consistent deterioration in the forecast precision in terms of CRPS in all model specifications. Increasing the window to 182 days partially mitigates this effect and yields modest improvements in CRPS of historical simulation relative to the unrestricted case; however, this improvement remains confined to configurations employing SVS. A~similar pattern of accuracy decreasing with decreasing window size is also observed for the na\"{i}ve benchmark results. 
Since the introduction of the window size constitutes an additional hyperparameter, increasing the complexity of the model without delivering consistent performance gains, we further consider only the unrestricted window case.

\begin{table}[]
\centering
\resizebox{0.7\textwidth}{!}{%
\begin{tabular}{cc|crrr|crrr|}
\cline{3-10}
\multicolumn{1}{l}{} &
  \multicolumn{1}{l|}{} &
  \multicolumn{4}{c|}{MAE} &
  \multicolumn{4}{c|}{CRPS} \\ \cline{3-10} 
\multicolumn{1}{l}{} &
  \multicolumn{1}{l|}{} &
  \multicolumn{2}{c|}{multi} &
  \multicolumn{2}{c|}{chain} &
  \multicolumn{2}{c|}{multi} &
  \multicolumn{2}{c|}{chain} \\ \hline
\multicolumn{1}{|c|}{model} &
  window &
  \multicolumn{1}{c|}{SVS} &
  \multicolumn{1}{c|}{---} &
  \multicolumn{1}{c|}{SVS} &
  \multicolumn{1}{c|}{---} &
  \multicolumn{1}{c|}{SVS} &
  \multicolumn{1}{c|}{---} &
  \multicolumn{1}{c|}{SVS} &
  \multicolumn{1}{c|}{---} \\ \hline
\multicolumn{1}{|c|}{} &
  28 &
  \multicolumn{1}{r|}{\cellcolor[HTML]{9CCF87}4.979} &
  \multicolumn{1}{r|}{\cellcolor[HTML]{9CCF87}4.979} &
  \multicolumn{1}{r|}{\cellcolor[HTML]{ACD58A}4.991} &
  \cellcolor[HTML]{ACD58A}4.991 &
  \multicolumn{1}{r|}{\cellcolor[HTML]{A7D389}2.089} &
  \multicolumn{1}{r|}{\cellcolor[HTML]{A7D389}2.090} &
  \multicolumn{1}{r|}{\cellcolor[HTML]{9ED087}2.071} &
  \cellcolor[HTML]{9ED087}2.071 \\ \cline{2-10} 
\multicolumn{1}{|c|}{} &
  182 &
  \multicolumn{1}{r|}{\cellcolor[HTML]{B9D98D}5.000} &
  \multicolumn{1}{r|}{\cellcolor[HTML]{81C781}4.960} &
  \multicolumn{1}{r|}{\cellcolor[HTML]{CBDE91}5.013} &
  \cellcolor[HTML]{A0D187}4.982 &
  \multicolumn{1}{r|}{\cellcolor[HTML]{70C27D}1.976} &
  \multicolumn{1}{r|}{\cellcolor[HTML]{95CD85}2.052} &
  \multicolumn{1}{r|}{\cellcolor[HTML]{72C27E}1.980} &
  \cellcolor[HTML]{8ACA83}2.030 \\ \cline{2-10} 
\multicolumn{1}{|c|}{\multirow{-3}{*}{historical}} &
  --- &
  \multicolumn{1}{r|}{\cellcolor[HTML]{D5E193}5.020} &
  \multicolumn{1}{r|}{\cellcolor[HTML]{7AC580}4.955} &
  \multicolumn{1}{r|}{\cellcolor[HTML]{F8EC9A}5.045} &
  \cellcolor[HTML]{A0D187}4.982 &
  \multicolumn{1}{r|}{\cellcolor[HTML]{79C47F}1.993} &
  \multicolumn{1}{r|}{\cellcolor[HTML]{94CD85}2.049} &
  \multicolumn{1}{r|}{\cellcolor[HTML]{9ACF86}2.063} &
  \cellcolor[HTML]{89C983}2.026 \\ \hline
\multicolumn{1}{|c|}{} &
  28 &
  \multicolumn{1}{r|}{\cellcolor[HTML]{67BF7B}4.941} &
  \multicolumn{1}{r|}{\cellcolor[HTML]{67BF7B}4.941} &
  \multicolumn{1}{r|}{\cellcolor[HTML]{80C781}4.959} &
  \cellcolor[HTML]{7EC680}4.958 &
  \multicolumn{1}{r|}{\cellcolor[HTML]{FFEF9C}2.271} &
  \multicolumn{1}{r|}{\cellcolor[HTML]{FDEE9B}2.268} &
  \multicolumn{1}{r|}{\cellcolor[HTML]{FEEE9B}2.269} &
  \cellcolor[HTML]{FDEE9B}2.267 \\ \cline{2-10} 
\multicolumn{1}{|c|}{} &
  182 &
  \multicolumn{1}{r|}{\cellcolor[HTML]{6EC17D}4.946} &
  \multicolumn{1}{r|}{\cellcolor[HTML]{64BE7B}4.939} &
  \multicolumn{1}{r|}{\cellcolor[HTML]{80C781}4.959} &
  \cellcolor[HTML]{7AC580}4.955 &
  \multicolumn{1}{r|}{\cellcolor[HTML]{FAED9A}2.261} &
  \multicolumn{1}{r|}{\cellcolor[HTML]{F4EB99}2.250} &
  \multicolumn{1}{r|}{\cellcolor[HTML]{F3EB99}2.248} &
  \cellcolor[HTML]{F3EB99}2.248 \\ \cline{2-10} 
\multicolumn{1}{|c|}{\multirow{-3}{*}{fundamental}} &
  --- &
  \multicolumn{1}{r|}{\cellcolor[HTML]{6CC17D}4.945} &
  \multicolumn{1}{r|}{\cellcolor[HTML]{63BE7B}4.938} &
  \multicolumn{1}{r|}{\cellcolor[HTML]{8ACA83}4.966} &
  \cellcolor[HTML]{79C57F}4.954 &
  \multicolumn{1}{r|}{\cellcolor[HTML]{F9ED9A}2.260} &
  \multicolumn{1}{r|}{\cellcolor[HTML]{F3EB99}2.247} &
  \multicolumn{1}{r|}{\cellcolor[HTML]{F1EA99}2.242} &
  \cellcolor[HTML]{F2EB99}2.245 \\ \hline
\multicolumn{1}{|c|}{} &
  28 &
  \multicolumn{4}{c|}{\cellcolor[HTML]{FFEF9C}5.050} &
  \multicolumn{4}{c|}{\cellcolor[HTML]{7AC57F}1.995} \\ \cline{2-10} 
\multicolumn{1}{|c|}{} &
  182 &
  \multicolumn{4}{c|}{\cellcolor[HTML]{ACD58A}4.991} &
  \multicolumn{4}{c|}{\cellcolor[HTML]{65BE7B}1.953} \\ \cline{2-10} 
\multicolumn{1}{|c|}{\multirow{-3}{*}{na\"{i}ve}} &
  --- &
  \multicolumn{4}{c|}{\cellcolor[HTML]{A0D187}4.982} &
  \multicolumn{4}{c|}{\cellcolor[HTML]{63BE7B}1.947} \\ \hline
\end{tabular}%
}
\caption{Forecast accuracy of historical simulation, fundamental scenarios, and the naïve benchmark. Results are presented for both multi-output and regressor chain approaches. Scenarios are sampled from the full training set and restricted windows of 28 or 182 days. MAE \eqref{eq:MAE} and CRPS \eqref{eq:CRPS} metrics are evaluated over the out-of-sample test period.}
\label{tab:CRPS_MAE_results}
\end{table}

Next, we evaluate the relative performance of multi-output versus regressor chain approaches. For historical simulation without SVS, the chain regressor achieves a~marginal improvement of approximately 1\% in CRPS compared to the multi-output approach. 
 However, this advantage disappears if SVS is applied. Consequently, the effect of chain regression on probabilistic accuracy (CRPS) remains inconclusive. Regarding point forecasts, the chain-based models consistently yield higher MAE values, likely due to the propagation of estimation errors inherent in sequential forecasting. Since introducing sequential forecast dependencies does not improve accuracy, we further use only the multi-output specification. It offers superior point forecasts and comparable probabilistic performance while maintaining a more parsimonious structure with fewer structural assumptions.

We exclude both restricted sampling windows and regressor chain from further analysis and proceed by comparing the performance of historical, fundamental, and na\"{i}ve benchmark scenarios. Regarding the accuracy of point path forecasts, the fundamental scenarios yield the lowest MAE in all specifications considered. In the SVS case, the relative improvement amounts to 1.5\% with respect to the historical simulation and 0.7\% with respect to the na\"{i}ve benchmark. In the absence of SVS, the corresponding improvements are 0.3\% and 0.9\%, respectively. In contrast, the CRPS associated with fundamental scenarios is markedly higher than that of the alternative approaches. The best CRPS is achieved by the na\"{i}ve benchmark, while the second-best performance is obtained by the historical simulation with SVS, whose CRPS is 2.4\% higher than that of the benchmark.

\begin{figure}[!h]
    \centering
    \includegraphics[width=\textwidth]{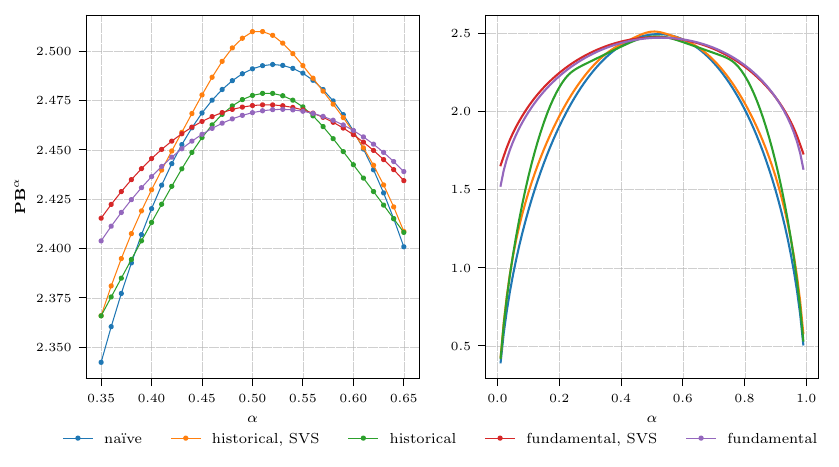}
    \caption{Pinball score (PB) averaged over days in test window, deliveries and path steps for different forecasting approaches. The forecasts are calculated using  unrestricted window and multi-output regression. The right panel represents the resulting pinball score aggregation over all 99 percentiles, while on the left panel a~zoomed-in center of the distribution is presented.}
    \label{fig:global_pinball}
\end{figure}

To further assess the forecast accuracy of the best-performing models, as well as the impact of the SVS procedure on the ensemble, in Figure \ref{fig:global_pinball} we plot the corresponding pinball scores for 99 quantiles, each averaged across all delivery periods and days in the test window. For  historical simulation the application of SVS leads to an increase in pinball score in the central region of the distribution, while improving accuracy in the tails, specifically below the 35th and above the 65th percentiles. This pattern suggests that the SVS procedure, together with the Wasserstein-based thresholding, effectively removes redundant scenarios from the historical sample, thereby enhancing the representation of mildly extreme prices. In contrast, this effect is not observed for the fundamental scenarios, where the application of SVS does not lead to a reduction in the overall CRPS. Moreover, although fundamental scenarios exhibit superior performance in the center of the distribution, this advantage diminishes rapidly outside this region. As a~result, they yield higher pinball score values, particularly in the lower and upper tails, i.e. below the 20th and above the 80th percentiles.

Finally, we examine the impact of representative scenario selection on the effective sample size of forecast scenarios, as reported in Table \ref{tab:scenarios_number}. The dispersion of the number of trajectories selected by the SVS procedure across the expanding window is markedly smaller than the variability induced by the natural expansion of the sample. For example, in the case of historical simulation with SVS the ratio of the first quartile (Q1) to the median amounts to 84\%, compared to only 70\% in the absence of SVS. When considered jointly with the observed improvements in CRPS, this result suggests that the application of SVS to historical simulation effectively isolates a relatively small but economically diverse subset of scenarios. Empirically, in 75\% of cases, fewer than 100 trajectories are sufficient to approximate the full distribution in terms of the Wasserstein distance. For fundamental scenarios the relevant manifold appears to collapse more sharply, which is reflected in the substantially lower number of selected trajectories. However, as shown in Figure \ref{fig:global_pinball} this reduction compromises the model's probabilistic performance only marginally. The low scenario count indicates that renewable generation and consumption forecast errors induce only limited perturbations in the cSVR outputs, resulting in highly similar price trajectories. Consequently, despite their theoretical appeal, fundamental corrections do not translate into improved CRPS performance, as also observed in Table \ref{tab:CRPS_MAE_results}, due to the concentration of informational content. By contrast, historical trajectories retain both structural and micro-structural heterogeneity, such as varying liquidity conditions, ramping episodes, and imbalance states. Hence, the historical simulation supports a~richer and more informative scenario representation.

\begin{table}[!h]
\centering
\resizebox{0.45\textwidth}{!}{%
\begin{tabular}{c|r|r|r|}
\cline{2-4}
 & \multicolumn{1}{c|}{---} & \multicolumn{1}{c|}{historical, SVS} & \multicolumn{1}{c|}{fundamental, SVS} \\ \hline
\multicolumn{1}{|c|}{mean}   & 456 & 88  & 28 \\ \hline
\multicolumn{1}{|c|}{Q1}     & 319 & 73  & 23 \\ \hline
\multicolumn{1}{|c|}{median} & 456 & 87  & 27 \\ \hline
\multicolumn{1}{|c|}{Q3}     & 592 & 101 & 33 \\ \hline
\end{tabular}%
}
\caption{Distribution of the number of scenarios across the entire expanding window scheme for different selection approaches and multi-output regression.}
\label{tab:scenarios_number}
\end{table}

\subsection{Trading strategies}
{We apply the ensembles of path forecasts derived using fundamental scenarios, historical simulation and the na\"{i}ve benchmark to evaluate trading strategies described in Section \ref{sec:trading_strategies}. Based on the forecast accuracy results, we restrict our analysis to the multi-output approach and scenarios sampled from non-restricted~windows.}
 
The {dynamically adapted} trading strategies depend on a set of hyperparameters governing tail sensitivity, temporal decay, and frequency of updating the strategy decisions. We select these parameter values via an exhaustive grid search on the calibration window, encompassing 182 days between 3rd July 2019 and 31st December 2019. The optimal configuration is fixed ex ante and subsequently evaluated on the out-of-sample period. This procedure is conducted separately for each strategy type and model combination. 

For the median-based strategies, the parameter grid consists of: shape parameter $p$, exponential decay parameter $\lambda$ and trust-threshold estimation method $\eta$. The considered values of these parameters are listed in Table \ref{tab:median_strategy_hyperparameters}.
\begin{table}[h]
\centering
\begin{adjustbox}{width=\textwidth}
\begin{tabular}{p{9cm} l}
\toprule
Description & Candidate values \\
\midrule

Shape parameter & 
$p \in \{0.1, 0.25, 0.5, 0.75, 1, 1.25, 1.5, 1.75, 2, 2.25, 2.5, 2.75, 3\}$ \\

Exponential decay parameter & 
$\lambda \in \{0, 0.05, 0.1, \dots, 0.5\}$ \\

Trust-threshold estimation method & 
$\eta \in \{3\sigma, \text{IQR}, \text{5--95 IPR}, \text{MAE}\}$ \\

\bottomrule
\end{tabular}
\end{adjustbox}
\caption{Grid of hyperparameters considered in the optimization procedure for dynamically adapted median-based strategies. }
\label{tab:median_strategy_hyperparameters}
\end{table}
For the band-based strategies, the grid additionally includes the SCP. Thus, to keep the grid search computational time feasible, we limit the $p$, $\lambda$ and $\eta$ values to the best performing values found in the median grid search. We acknowledge that this is a~simplification, but, due to the broad range of values selected as optimal for different models under the median strategy, it should not impact the quality of grid search significantly. The final parameter space for band-based strategies is given in Table \ref{tab:bands_strategy_hyperparameters}.

\begin{table}[h]
\centering
\begin{adjustbox}{width=0.8\textwidth}
\begin{tabular}{p{9cm} l}
\toprule
Description & Candidate values \\
\midrule

Simultaneous coverage probability & 
$\mathrm{SCP} \in \{0.05, 0.10, \dots, 0.95\}$ \\

Shape parameter & 
$p \in \{0.5, 0.75, 1.0, 1.25, 1.75, 2.0\}$ \\

Exponential decay parameter & 
$\lambda \in \{0.05, 0.1, 0.2, 0.35, 0.4, 0.5\}$ \\

Trust-threshold estimation method & 
$\eta \in \{\text{IQR}, \text{5--95 IPR}\}$ \\

\bottomrule
\end{tabular}
\end{adjustbox}
\caption{Grid of hyperparameters considered in the optimization procedure for dynamically adapted band-based strategies.}
\label{tab:bands_strategy_hyperparameters}
\end{table}

For each parameter configuration $\theta$ in the corresponding Cartesian product, we compute the profit and loss (PnL) sequence 
 over the calibration window. For the seller, the profit equals the transaction price
\begin{equation}
    \pi_{d,T}(\theta) = P_{d,T}(u),
\end{equation}
where $u$ is the path step at which the energy was sold. 
For the spread trader, the profit is calculated as 
the difference between the prices at the exit ($u_1$) and entry ($u_0$) times multiplied by the direction of trade $s_{d,T}$
\begin{equation}
    \pi_{d,T}(\theta) = s_{d,T}\left( P_{d,T}(u_1) - P_{d,T}(u_0) \right),
\end{equation}
where $s_{d,T}$ is equal to 1 or -1, for long or short position, respectively. Hyperparameter selection is then based on the minimization of the Sortino ratio, i.e. the ratio of the total profit to the downside risk. 
{The total profit of trading strategy for parameter configuration $\theta$, $\Pi(\theta)$, is defined as the sum of profits for all deliveries and days in the calibration window}
\begin{equation}
    \Pi(\theta) = \sum_{T=1}^{N}\sum_{d=1}^{96} \pi_{d,T}(\theta).
\label{eq:total_profit}
\end{equation}
{For the downside risk measure, we use the semi-standard deviation}
\begin{equation}
    \sigma_{-}(\theta) = \sqrt{
        \frac{1}{96\cdot N}
        \sum_{T=1}^{N}\sum_{d=1}^{96}
        \bigl(\min\{\pi_{d,T}(\theta),\mu\}\bigr)^2
        },
    \label{eq:downside_risk}
\end{equation}
where $\mu$ is set to $0$ for the spread trader and the mean of the PnL sequence for the seller. The Sortino ratio, used in a~similar context, e.g., by \cite{sortino_ratio_in_energy_trading_context}, is then formally defined as
\begin{equation}
\mathrm{Sortino}(\theta)
=
\frac{
 \Pi(\theta)
}{
\sigma_{-}(\theta)
}.
\end{equation}

The grid search results reported in the Appendix, Table~\ref{tab:strrategies_grid_search_results}, reveal the emergence of two distinct regimes driven by the interaction between the tail parameter $p$~and the intra-path memory parameter $\lambda$. For lower values of $p$, the kernel induces a~gradual decay of the scenario weights, leading to smoother transitions between the consecutive median forecast paths. This regime is systematically preferred in the spread trader's case, whose payoff depends primarily on relative price differentials rather than absolute price levels. In this setting, higher values of $\lambda$ are selected, prioritizing the most recent path steps and thus allowing corrective adjustments in the case of extreme price changes, even in the presence of low $p$. In contrast, higher values of $p$ concentrate the weight mass around scenarios closest to the realized price, resulting in more reactive and occasionally abrupt adjustments of the median paths. This regime is favored in the seller's case, for whom accurate tracking of absolute price levels is critical. Consistent with this objective, lower values of $\lambda$ are selected, reducing sensitivity to intra-path fluctuations and limiting the frequency of strategy updates. The distinction between these regimes is further reinforced by the choice of decision thresholds: the spread trader's strategy predominantly adopts wider bands (e.g., 5--95 IPR), promoting stability and robustness to noise, whereas the seller's strategy relies on narrower bands (typically IQR), enabling more frequent and responsive adjustments.

A~notable interaction between the simultaneous coverage probability (SCP) and the risk preference specification emerges for the seller. In particular, the SCP associated with risk-seeking configurations consistently attains its lower bound in grid search, i.e. $5\%$, while the SCP for risk-averse configurations is pushed towards its upper bound, i.e. $95\%$. This pattern indicates that the grid search systematically compensates for the imposed risk preferences through opposing adjustments in the probabilistic coverage. Specifically, risk-averse strategies are effectively made less conservative by expanding the lower tail coverage, whereas risk-seeking strategies are tempered by restricting the upper tail exposure. Such behavior is caused by the underlying grid search loss function, namely the Sortino ratio, which accounts for both, risk and profit.

\subsubsection{Results}

\begin{table}[]
\centering
\resizebox{0.85\textwidth}{!}{%
\begin{tabular}{|ccccc|rr|rr|rr|}
\hline
\multicolumn{5}{|c|}{strategy specification} &
  \multicolumn{2}{c|}{dynamic   kernel} &
  \multicolumn{2}{c|}{dynamic MAE} &
  \multicolumn{2}{c|}{static} \\ \hline
\multicolumn{1}{|c|}{agent} &
  \multicolumn{1}{c|}{strategy} &
  \multicolumn{1}{c|}{\begin{tabular}[c]{@{}c@{}}band\\ type\end{tabular}} &
  \multicolumn{1}{c|}{\begin{tabular}[c]{@{}c@{}}scenario\\ selection\end{tabular}} &
  model &
  \multicolumn{1}{c|}{$\Pi$} &
  \multicolumn{1}{c|}{$\sigma_{-}$} &
  \multicolumn{1}{c|}{$\Pi$} &
  \multicolumn{1}{c|}{$\sigma_{-}$} &
  \multicolumn{1}{c|}{$\Pi$} &
  \multicolumn{1}{c|}{$\sigma_{-}$} \\ \hline
\multicolumn{1}{|c|}{} &
  \multicolumn{1}{c|}{} &
  \multicolumn{1}{c|}{} &
  \multicolumn{1}{c|}{} &
  na\"{i}ve &
  \multicolumn{1}{r|}{\cellcolor[HTML]{F3EC9A}12837} &
  \cellcolor[HTML]{63BE7B}7.33 &
  \multicolumn{1}{r|}{\cellcolor[HTML]{FFEF9C}8022} &
  \cellcolor[HTML]{9CCF87}11.49 &
  \multicolumn{1}{r|}{\cellcolor[HTML]{FFEF9C}7794} &
  \cellcolor[HTML]{FFEF9C}18.70 \\ \cline{5-11} 
\multicolumn{1}{|c|}{} &
  \multicolumn{1}{c|}{} &
  \multicolumn{1}{c|}{} &
  \multicolumn{1}{c|}{} &
  historical &
  \multicolumn{1}{r|}{\cellcolor[HTML]{A5D389}44300} &
  \cellcolor[HTML]{65BE7B}7.49 &
  \multicolumn{1}{r|}{\cellcolor[HTML]{B5D88D}37719} &
  \cellcolor[HTML]{93CD85}10.83 &
  \multicolumn{1}{r|}{\cellcolor[HTML]{95CE86}50562} &
  \cellcolor[HTML]{89CA83}10.12 \\ \cline{5-11} 
\multicolumn{1}{|c|}{} &
  \multicolumn{1}{c|}{} &
  \multicolumn{1}{c|}{} &
  \multicolumn{1}{c|}{\multirow{-3}{*}{---}} &
  fundamental &
  \multicolumn{1}{r|}{\cellcolor[HTML]{74C47F}63717} &
  \cellcolor[HTML]{B9D98D}13.63 &
  \multicolumn{1}{r|}{\cellcolor[HTML]{79C580}61764} &
  \cellcolor[HTML]{BCDA8D}13.83 &
  \multicolumn{1}{r|}{\cellcolor[HTML]{77C580}62860} &
  \cellcolor[HTML]{B9D98D}13.62 \\ \cline{4-11} 
\multicolumn{1}{|c|}{} &
  \multicolumn{1}{c|}{} &
  \multicolumn{1}{c|}{} &
  \multicolumn{1}{c|}{SVS} &
  historical &
  \multicolumn{1}{r|}{\cellcolor[HTML]{B0D78C}39726} &
  \cellcolor[HTML]{BFDB8E}14.08 &
  \multicolumn{1}{r|}{\cellcolor[HTML]{9ED188}46836} &
  \cellcolor[HTML]{B2D68B}13.12 &
  \multicolumn{1}{r|}{\cellcolor[HTML]{A5D389}44057} &
  \cellcolor[HTML]{A9D489}12.48 \\ \cline{4-11} 
\multicolumn{1}{|c|}{} &
  \multicolumn{1}{c|}{} &
  \multicolumn{1}{c|}{\multirow{-5}{*}{\begin{tabular}[c]{@{}c@{}}risk\\ averse\end{tabular}}} &
  \multicolumn{1}{c|}{SVS} &
  fundamental &
  \multicolumn{1}{r|}{\cellcolor[HTML]{76C480}62928} &
  \cellcolor[HTML]{B9D98D}13.63 &
  \multicolumn{1}{r|}{\cellcolor[HTML]{83C882}57810} &
  \cellcolor[HTML]{B7D88C}13.47 &
  \multicolumn{1}{r|}{\cellcolor[HTML]{73C37F}64208} &
  \cellcolor[HTML]{B8D88D}13.54 \\ \cline{3-11} 
\multicolumn{1}{|c|}{} &
  \multicolumn{1}{c|}{} &
  \multicolumn{1}{c|}{} &
  \multicolumn{1}{c|}{} &
  na\"{i}ve &
  \multicolumn{1}{r|}{\cellcolor[HTML]{E1E696}20217} &
  \cellcolor[HTML]{B9D98D}13.63 &
  \multicolumn{1}{r|}{\cellcolor[HTML]{E9E898}16852} &
  \cellcolor[HTML]{7BC580}9.10 &
  \multicolumn{1}{r|}{\cellcolor[HTML]{FDEF9C}8612} &
  \cellcolor[HTML]{99CF86}11.29 \\ \cline{5-11} 
\multicolumn{1}{|c|}{} &
  \multicolumn{1}{c|}{} &
  \multicolumn{1}{c|}{} &
  \multicolumn{1}{c|}{} &
  historical &
  \multicolumn{1}{r|}{\cellcolor[HTML]{86C983}56652} &
  \cellcolor[HTML]{A0D188}11.84 &
  \multicolumn{1}{r|}{\cellcolor[HTML]{96CE86}50373} &
  \cellcolor[HTML]{B3D78C}13.23 &
  \multicolumn{1}{r|}{\cellcolor[HTML]{9BD087}48259} &
  \cellcolor[HTML]{AAD48A}12.56 \\ \cline{5-11} 
\multicolumn{1}{|c|}{} &
  \multicolumn{1}{c|}{} &
  \multicolumn{1}{c|}{} &
  \multicolumn{1}{c|}{\multirow{-3}{*}{---}} &
  fundamental &
  \multicolumn{1}{r|}{\cellcolor[HTML]{6BC17D}67590} &
  \cellcolor[HTML]{BEDA8E}14.02 &
  \multicolumn{1}{r|}{\cellcolor[HTML]{7DC781}60198} &
  \cellcolor[HTML]{B5D78C}13.31 &
  \multicolumn{1}{r|}{\cellcolor[HTML]{90CC85}52836} &
  \cellcolor[HTML]{AFD68B}12.92 \\ \cline{4-11} 
\multicolumn{1}{|c|}{} &
  \multicolumn{1}{c|}{} &
  \multicolumn{1}{c|}{} &
  \multicolumn{1}{c|}{SVS} &
  historical &
  \multicolumn{1}{r|}{\cellcolor[HTML]{9ED188}47187} &
  \cellcolor[HTML]{9CD087}11.51 &
  \multicolumn{1}{r|}{\cellcolor[HTML]{9DD188}47374} &
  \cellcolor[HTML]{ADD58A}12.78 &
  \multicolumn{1}{r|}{\cellcolor[HTML]{9BD087}48398} &
  \cellcolor[HTML]{BBD98D}13.74 \\ \cline{4-11} 
\multicolumn{1}{|c|}{} &
  \multicolumn{1}{c|}{\multirow{-10}{*}{bands}} &
  \multicolumn{1}{c|}{\multirow{-5}{*}{\begin{tabular}[c]{@{}c@{}}risk\\ seeking\end{tabular}}} &
  \multicolumn{1}{c|}{SVS} &
  fundamental &
  \multicolumn{1}{r|}{\cellcolor[HTML]{76C47F}63205} &
  \cellcolor[HTML]{BCDA8D}13.84 &
  \multicolumn{1}{r|}{\cellcolor[HTML]{76C47F}62966} &
  \cellcolor[HTML]{BEDA8E}14.00 &
  \multicolumn{1}{r|}{\cellcolor[HTML]{7CC681}60795} &
  \cellcolor[HTML]{B8D98D}13.59 \\ \cline{2-11} 
\multicolumn{1}{|c|}{} &
  \multicolumn{1}{c|}{} &
  \multicolumn{1}{c|}{} &
  \multicolumn{1}{c|}{} &
  na\"{i}ve &
  \multicolumn{1}{r|}{\cellcolor[HTML]{B0D78C}39782} &
  \cellcolor[HTML]{EFEA98}17.59 &
  \multicolumn{1}{r|}{\cellcolor[HTML]{BEDB8F}34310} &
  \cellcolor[HTML]{EDE998}17.44 &
  \multicolumn{1}{r|}{\cellcolor[HTML]{AAD58A}42144} &
  \cellcolor[HTML]{EEE998}17.50 \\ \cline{5-11} 
\multicolumn{1}{|c|}{} &
  \multicolumn{1}{c|}{} &
  \multicolumn{1}{c|}{} &
  \multicolumn{1}{c|}{} &
  historical &
  \multicolumn{1}{r|}{\cellcolor[HTML]{64BF7C}70273} &
  \cellcolor[HTML]{AAD48A}12.52 &
  \multicolumn{1}{r|}{\cellcolor[HTML]{6BC17D}67733} &
  \cellcolor[HTML]{BEDA8E}13.98 &
  \multicolumn{1}{r|}{\cellcolor[HTML]{68C07D}68559} &
  \cellcolor[HTML]{B8D88D}13.54 \\ \cline{5-11} 
\multicolumn{1}{|c|}{} &
  \multicolumn{1}{c|}{} &
  \multicolumn{1}{c|}{} &
  \multicolumn{1}{c|}{\multirow{-3}{*}{---}} &
  fundamental &
  \multicolumn{1}{r|}{\cellcolor[HTML]{65BF7C}69939} &
  \cellcolor[HTML]{C3DC8F}14.34 &
  \multicolumn{1}{r|}{\cellcolor[HTML]{6CC17D}67041} &
  \cellcolor[HTML]{C4DC8F}14.45 &
  \multicolumn{1}{r|}{\cellcolor[HTML]{6DC27E}66696} &
  \cellcolor[HTML]{C2DB8F}14.27 \\ \cline{4-11} 
\multicolumn{1}{|c|}{} &
  \multicolumn{1}{c|}{} &
  \multicolumn{1}{c|}{} &
  \multicolumn{1}{c|}{SVS} &
  historical &
  \multicolumn{1}{r|}{\cellcolor[HTML]{76C47F}63008} &
  \cellcolor[HTML]{BFDB8E}14.09 &
  \multicolumn{1}{r|}{\cellcolor[HTML]{7DC781}60299} &
  \cellcolor[HTML]{BEDA8E}14.00 &
  \multicolumn{1}{r|}{\cellcolor[HTML]{78C580}62334} &
  \cellcolor[HTML]{BAD98D}13.74 \\ \cline{4-11} 
\multicolumn{1}{|c|}{\multirow{-15}{*}{\begin{tabular}[c]{@{}c@{}}spread\\ trader\end{tabular}}} &
  \multicolumn{1}{c|}{\multirow{-5}{*}{median}} &
  \multicolumn{1}{c|}{\multirow{-5}{*}{---}} &
  \multicolumn{1}{c|}{SVS} &
  fundamental &
  \multicolumn{1}{r|}{\cellcolor[HTML]{63BE7B}70550} &
  \cellcolor[HTML]{C5DC8F}14.51 &
  \multicolumn{1}{r|}{\cellcolor[HTML]{71C37E}65112} &
  \cellcolor[HTML]{C9DE90}14.79 &
  \multicolumn{1}{r|}{\cellcolor[HTML]{6AC17D}67986} &
  \cellcolor[HTML]{C0DB8E}14.14 \\ \hline
\multicolumn{1}{|c|}{} &
  \multicolumn{1}{c|}{} &
  \multicolumn{1}{c|}{} &
  \multicolumn{1}{c|}{} &
  na\"{i}ve &
  \multicolumn{1}{r|}{\cellcolor[HTML]{B6D98D}1138318} &
  \cellcolor[HTML]{68BF7C}15.30 &
  \multicolumn{1}{r|}{\cellcolor[HTML]{C0DC8F}1132367} &
  \cellcolor[HTML]{69BF7C}15.31 &
  \multicolumn{1}{r|}{\cellcolor[HTML]{FFEF9C}1093037} &
  \cellcolor[HTML]{74C37E}15.45 \\ \cline{5-11} 
\multicolumn{1}{|c|}{} &
  \multicolumn{1}{c|}{} &
  \multicolumn{1}{c|}{} &
  \multicolumn{1}{c|}{} &
  historical &
  \multicolumn{1}{r|}{\cellcolor[HTML]{85C983}1168896} &
  \cellcolor[HTML]{76C47F}15.47 &
  \multicolumn{1}{r|}{\cellcolor[HTML]{8ECC84}1163375} &
  \cellcolor[HTML]{79C47F}15.51 &
  \multicolumn{1}{r|}{\cellcolor[HTML]{F5EC9A}1099534} &
  \cellcolor[HTML]{81C781}15.61 \\ \cline{5-11} 
\multicolumn{1}{|c|}{} &
  \multicolumn{1}{c|}{} &
  \multicolumn{1}{c|}{} &
  \multicolumn{1}{c|}{\multirow{-3}{*}{---}} &
  fundamental &
  \multicolumn{1}{r|}{\cellcolor[HTML]{97CF86}1157800} &
  \cellcolor[HTML]{92CC85}15.83 &
  \multicolumn{1}{r|}{\cellcolor[HTML]{9AD087}1155768} &
  \cellcolor[HTML]{8ECB84}15.77 &
  \multicolumn{1}{r|}{\cellcolor[HTML]{DEE595}1113855} &
  \cellcolor[HTML]{98CE86}15.90 \\ \cline{4-11} 
\multicolumn{1}{|c|}{} &
  \multicolumn{1}{c|}{} &
  \multicolumn{1}{c|}{} &
  \multicolumn{1}{c|}{SVS} &
  historical &
  \multicolumn{1}{r|}{\cellcolor[HTML]{9AD087}1155972} &
  \cellcolor[HTML]{6AC07C}15.33 &
  \multicolumn{1}{r|}{\cellcolor[HTML]{A8D48A}1147300} &
  \cellcolor[HTML]{63BE7B}15.23 &
  \multicolumn{1}{r|}{\cellcolor[HTML]{FCEE9C}1095379} &
  \cellcolor[HTML]{7BC580}15.54 \\ \cline{4-11} 
\multicolumn{1}{|c|}{} &
  \multicolumn{1}{c|}{} &
  \multicolumn{1}{c|}{\multirow{-5}{*}{\begin{tabular}[c]{@{}c@{}}risk\\ averse\end{tabular}}} &
  \multicolumn{1}{c|}{SVS} &
  fundamental &
  \multicolumn{1}{r|}{\cellcolor[HTML]{81C882}1171356} &
  \cellcolor[HTML]{B0D68B}16.20 &
  \multicolumn{1}{r|}{\cellcolor[HTML]{8ACB84}1165788} &
  \cellcolor[HTML]{A9D489}16.11 &
  \multicolumn{1}{r|}{\cellcolor[HTML]{D2E193}1121440} &
  \cellcolor[HTML]{ADD58A}16.16 \\ \cline{3-11} 
\multicolumn{1}{|c|}{} &
  \multicolumn{1}{c|}{} &
  \multicolumn{1}{c|}{} &
  \multicolumn{1}{c|}{} &
  na\"{i}ve &
  \multicolumn{1}{r|}{\cellcolor[HTML]{ADD68B}1143933} &
  \cellcolor[HTML]{63BE7B}15.23 &
  \multicolumn{1}{r|}{\cellcolor[HTML]{B2D78C}1140721} &
  \cellcolor[HTML]{63BE7B}15.23 &
  \multicolumn{1}{r|}{\cellcolor[HTML]{FCEE9C}1095066} &
  \cellcolor[HTML]{71C27E}15.41 \\ \cline{5-11} 
\multicolumn{1}{|c|}{} &
  \multicolumn{1}{c|}{} &
  \multicolumn{1}{c|}{} &
  \multicolumn{1}{c|}{} &
  historical &
  \multicolumn{1}{r|}{\cellcolor[HTML]{65BF7C}1188845} &
  \cellcolor[HTML]{9DD087}15.96 &
  \multicolumn{1}{r|}{\cellcolor[HTML]{63BE7B}1189507} &
  \cellcolor[HTML]{A1D188}16.02 &
  \multicolumn{1}{r|}{\cellcolor[HTML]{DAE495}1115943} &
  \cellcolor[HTML]{A7D389}16.09 \\ \cline{5-11} 
\multicolumn{1}{|c|}{} &
  \multicolumn{1}{c|}{} &
  \multicolumn{1}{c|}{} &
  \multicolumn{1}{c|}{\multirow{-3}{*}{---}} &
  fundamental &
  \multicolumn{1}{r|}{\cellcolor[HTML]{7BC681}1174691} &
  \cellcolor[HTML]{AED58A}16.17 &
  \multicolumn{1}{r|}{\cellcolor[HTML]{82C882}1170611} &
  \cellcolor[HTML]{A9D48A}16.12 &
  \multicolumn{1}{r|}{\cellcolor[HTML]{D6E394}1118611} &
  \cellcolor[HTML]{ADD58A}16.17 \\ \cline{4-11} 
\multicolumn{1}{|c|}{} &
  \multicolumn{1}{c|}{} &
  \multicolumn{1}{c|}{} &
  \multicolumn{1}{c|}{SVS} &
  historical &
  \multicolumn{1}{r|}{\cellcolor[HTML]{91CD85}1161660} &
  \cellcolor[HTML]{7FC781}15.59 &
  \multicolumn{1}{r|}{\cellcolor[HTML]{9FD188}1152816} &
  \cellcolor[HTML]{74C37E}15.45 &
  \multicolumn{1}{r|}{\cellcolor[HTML]{F1EB9A}1101780} &
  \cellcolor[HTML]{8BCA83}15.73 \\ \cline{4-11} 
\multicolumn{1}{|c|}{} &
  \multicolumn{1}{c|}{\multirow{-10}{*}{bands}} &
  \multicolumn{1}{c|}{\multirow{-5}{*}{\begin{tabular}[c]{@{}c@{}}risk\\ seeking\end{tabular}}} &
  \multicolumn{1}{c|}{SVS} &
  fundamental &
  \multicolumn{1}{r|}{\cellcolor[HTML]{85C983}1168879} &
  \cellcolor[HTML]{A9D489}16.11 &
  \multicolumn{1}{r|}{\cellcolor[HTML]{8CCB84}1164575} &
  \cellcolor[HTML]{A6D389}16.08 &
  \multicolumn{1}{r|}{\cellcolor[HTML]{CDE092}1124469} &
  \cellcolor[HTML]{BFDB8E}16.40 \\ \cline{2-11} 
\multicolumn{1}{|c|}{} &
  \multicolumn{1}{c|}{} &
  \multicolumn{1}{c|}{} &
  \multicolumn{1}{c|}{} &
  na\"{i}ve &
  \multicolumn{1}{r|}{\cellcolor[HTML]{D3E193}1120621} &
  \cellcolor[HTML]{E8E797}16.91 &
  \multicolumn{1}{r|}{\cellcolor[HTML]{CEE092}1123627} &
  \cellcolor[HTML]{E6E796}16.88 &
  \multicolumn{1}{r|}{\cellcolor[HTML]{CBDF91}1125546} &
  \cellcolor[HTML]{E6E796}16.88 \\ \cline{5-11} 
\multicolumn{1}{|c|}{} &
  \multicolumn{1}{c|}{} &
  \multicolumn{1}{c|}{} &
  \multicolumn{1}{c|}{} &
  historical &
  \multicolumn{1}{r|}{\cellcolor[HTML]{BEDB8F}1133712} &
  \cellcolor[HTML]{ECE998}16.96 &
  \multicolumn{1}{r|}{\cellcolor[HTML]{B8D98D}1137403} &
  \cellcolor[HTML]{F1EA99}17.03 &
  \multicolumn{1}{r|}{\cellcolor[HTML]{B6D98D}1138255} &
  \cellcolor[HTML]{E9E897}16.92 \\ \cline{5-11} 
\multicolumn{1}{|c|}{} &
  \multicolumn{1}{c|}{} &
  \multicolumn{1}{c|}{} &
  \multicolumn{1}{c|}{\multirow{-3}{*}{---}} &
  fundamental &
  \multicolumn{1}{r|}{\cellcolor[HTML]{B9D98E}1136657} &
  \cellcolor[HTML]{F0EA98}17.00 &
  \multicolumn{1}{r|}{\cellcolor[HTML]{B8D98D}1137235} &
  \cellcolor[HTML]{F9ED9A}17.12 &
  \multicolumn{1}{r|}{\cellcolor[HTML]{B8D98D}1137043} &
  \cellcolor[HTML]{EFEA98}17.00 \\ \cline{4-11} 
\multicolumn{1}{|c|}{} &
  \multicolumn{1}{c|}{} &
  \multicolumn{1}{c|}{} &
  \multicolumn{1}{c|}{SVS} &
  historical &
  \multicolumn{1}{r|}{\cellcolor[HTML]{CEE092}1123623} &
  \cellcolor[HTML]{DFE595}16.80 &
  \multicolumn{1}{r|}{\cellcolor[HTML]{BDDA8E}1134391} &
  \cellcolor[HTML]{F3EB99}17.05 &
  \multicolumn{1}{r|}{\cellcolor[HTML]{BBDA8E}1135129} &
  \cellcolor[HTML]{ECE998}16.96 \\ \cline{4-11} 
\multicolumn{1}{|c|}{\multirow{-15}{*}{seller}} &
  \multicolumn{1}{c|}{\multirow{-5}{*}{median}} &
  \multicolumn{1}{c|}{\multirow{-5}{*}{---}} &
  \multicolumn{1}{c|}{SVS} &
  fundamental &
  \multicolumn{1}{r|}{\cellcolor[HTML]{B8D98D}1137414} &
  \cellcolor[HTML]{F8EC9A}17.11 &
  \multicolumn{1}{r|}{\cellcolor[HTML]{BADA8E}1135998} &
  \cellcolor[HTML]{FFEF9C}17.19 &
  \multicolumn{1}{r|}{\cellcolor[HTML]{B7D98D}1138117} &
  \cellcolor[HTML]{F6EC9A}17.08 \\ \hline
\end{tabular}%
}
\caption{{Total} profit $\Pi$, \eqref{eq:total_profit}, and average downside risk per delivery $\sigma_-$, \eqref{eq:downside_risk}, for all strategies and model specifications considered in case study. Color scales are separate for spread trader and seller and for profit and downside risk. The greenest cells contain the best results in their category. Values are given in EUR.}
\label{tab:strategies_result}
\end{table}

The dynamic strategy parameters, calibrated via grid search in the calibration window, are subsequently used for the out-of-sample evaluation of each strategy configuration. Outcomes of each static strategy are compared to its dynamic kernel- and MAE-based counterparts. The resulting {total} profit and downside risk metrics {per delivery} are reported in Table~\ref{tab:strategies_result}. 

Let us first discuss the results for the spread trader. Across all strategy classes and model specifications, the na\"{i}ve benchmark consistently yields the lowest profit levels. Although it is associated with relatively low downside risk, this does not compensate for its substantially inferior profitability. This confirms that purely persistence-based forecasts are insufficient to extract intraday spread opportunities. 

Comparing strategy types for the spread trader, the median-based strategies generally outperform band-based strategies in terms of profit, but at the cost of elevated risk levels. Within the band-based class, a~systematic distinction emerges between risk-seeking and risk-averse configurations. Under the dynamic kernel specification, the risk-seeking bands consistently achieve higher profits than their risk-averse counterparts. In certain cases the risk-seeking bands also attain lower risk; notably under historical simulation with SVS filtering. This pattern is partially preserved under the dynamic MAE specification, but breaks down under the static specification, where risk-averse bands tend to dominate both in terms of profit and risk.

The role of scenario construction is also pronounced in the spread trader strategy results. Fundamental scenarios systematically yield higher profits than historical simulation in most configurations. The only notable exception occurs for median-based strategies without scenario selection, where historical simulation marginally outperforms fundamental scenarios in terms of profit. The systematically higher profit of fundamental scenarios is obtained at the expense of higher risk. This is consistent with their weaker tail performance, as documented in Figure~\ref{fig:global_pinball}, which leads to increased exposure to extreme price realizations.

Finally, dynamic kernel-based strategy adjustments generally improve performance across configurations for the spread trader. This improvement is visible especially for the weaker baseline models like the na\"{i}ve, where we see
a nearly 135\% of relative improvement in profit in the risk seeking bands setting. Aggregating across all spread trader configurations, the median {relative} improvement in profit over the static approach attributable to the dynamic kernel is approximately $2.5\%$, with a median increase in downside risk of only $0.5\%$. The corresponding mean effects are $15\%$ and $4\%$, respectively. These results indicate that dynamic kernel adjustment enhances profitability at a~relatively low cost in terms of risk, consistent with its design, which emphasizes responsiveness to recent price realizations over explicit volatility control. In contrast, dynamic MAE-based strategies perform worse than both kernel-based and static approaches, resulting in a~median profit decrease of $1.2\%$ and a~median increase in risk of $2\%$ over the static counterpart. This result indicates that purely error-driven adjustments are less effective in the spread trader setting.

Turning to the seller, many qualitative patterns remain consistent, albeit with notable differences in magnitude and relative importance. The na\"{i}ve specification again underperforms in terms of profit; however, the performance gap relative to other models is substantially smaller than in the spread trader's case. Dynamic kernel adjustments lead to an increase in~median profit of approximately $4\%$ and $3\%$ on average, while simultaneously reducing the downside risk by $0.8\%$ (median) and $0.6\%$ (mean), indicating that, in this setting, dynamic adaptation can improve both dimensions of performance. The dynamic MAE weighting exhibits similar, albeit slightly weaker, improvements with a median increase in profit of $3.7\%$ and a reduction in risk of $0.4\%$.

Consistently with the results for the spread trader, the fundamental scenarios for the seller generally outperform historical simulations in terms of profitability. However, the highest overall profit is achieved using risk-seeking bands applied to historical scenarios without SVS filtering. Furthermore, in contrast to the spread trader's framework, band-based strategies for the seller dominate median-based approaches in both return and risk. This disparity in risk is particularly pronounced, as median-based strategies exhibit substantially higher downside variability. Such findings underscore the importance of leveraging the full predictive distribution: strategies relying exclusively on median forecasts disregard critical tail information, leading to inferior risk-adjusted performance in level-oriented trading contexts.

\begin{table}[]
\centering
\resizebox{0.5\textwidth}{!}{%
\begin{tabular}{|c|c|r|r|}
\hline
agent & \begin{tabular}[c]{@{}c@{}}strategy\\ specification\end{tabular} & \multicolumn{1}{c|}{$\Pi$} & \multicolumn{1}{c|}{$\sigma_{-}$} \\ \hline
\multirow{5}{*}{spread trader} & na\"{i}ve\textsubscript{first} & -30953  & 21.41 \\ \cline{2-4} 
                               & na\"{i}ve\textsubscript{last}  & 30953   & 11.94 \\ \cline{2-4} 
                               & crystal ball                   & 710939  & 0.00  \\ \cline{2-4} 
                               & opposite crystal ball          & -710939 & 38.13 \\ \cline{2-4} 
                               & best model                     & 70273   & 12.52 \\ \hline
\multirow{5}{*}{seller}        & na\"{i}ve\textsubscript{first} & 1093474 & 15.33 \\ \cline{2-4} 
                               & na\"{i}ve\textsubscript{last}  & 1124427 & 18.54 \\ \cline{2-4} 
                               & crystal ball                   & 1476217 & 15.64 \\ \cline{2-4} 
                               & opposite crystal ball          & 765277  & 19.79 \\ \cline{2-4} 
                               & best model                     & 1188845 & 15.96 \\ \hline
\end{tabular}%
}
\caption{The total profit $\Pi$, \eqref{eq:total_profit}, and average downside risk per delivery $\sigma_-$, \eqref{eq:downside_risk}, obtained with the na\"{i}ve and crystal ball benchmarks as well as the best performing models specifications. The latter are median strategy for the spread trader and the risk seeking bands for the seller, both based on historical scenarios. Values are given in EUR.}
\label{tab:crystal_ball_and_naive_strategies}
\end{table}

We summarize the performance of the proposed trading strategies by identifying the best-performing specifications and benchmarking them against the na\"{i}ve strategies and the crystal-ball reference reported in Table~\ref{tab:crystal_ball_and_naive_strategies}. For the seller, the na\"{i}ve benchmarks correspond to selling at the first (na\"{i}ve\textsubscript{first}) and last (na\"{i}ve\textsubscript{last}) step of the trajectory. For the spread trader, the first benchmark involves selling at the initial step and buying at the final step, whereas the second corresponds to the reverse position.

For the spread trader, the overall highest profit is achieved by the dynamic kernel median strategy applied to fundamental scenarios with SVS, yielding 70550 EUR with a downside risk {per delivery} of 14.51 EUR. However, when accounting jointly for profit and risk, the dynamic kernel median strategy based on historical simulation without SVS emerges as the most favorable specification, attaining 70273 EUR in {total} profit with a substantially lower downside risk of 12.52 EUR. Both configurations outperform the na\"{i}ve\textsubscript{last} benchmark by a factor of approximately 2.3, indicating that model-based forecasting substantially enhances the ability to exploit intra-path spread opportunities beyond the systematic upward drift observed along the trading trajectory. Notably, the downside risk of the best-performing specification remains comparable to that of the na\"{i}ve benchmark (11.94 EUR), showing that profitability gains are not driven by disproportionate increases in risk exposure. Compared to the crystal-ball benchmark, the best-performing spread trading strategies capture approximately 10\% of the attainable profit. This gap highlights the limitations of ensemble forecasts in fully recovering the ex post optimal trading trajectory, while simultaneously indicating scope for further improvements in scenario generation and decision rules.

Turning to the seller, the strongest performance is obtained by dynamic risk-seeking band strategies based on historical simulation without SVS. These configurations dominate their counterparts based on fundamental scenarios with SVS in both profit and risk dimensions. Moreover, the performance of the dynamic kernel and dynamic MAE specifications is broadly comparable, suggesting that for level-based strategies, frequent update plays a~more prominent role than the specific form of distributional reweighting. The best-performing configurations outperform the na\"{i}ve\textsubscript{last} benchmark, increasing profit by approximately 6\% while simultaneously reducing downside risk by around 14\%. In terms of attainable gains, the best seller strategies achieve approximately 60\% of the feasible performance range, measured as by \cite{SERAFIN2022106125} by normalizing the strategy profit between the opposite crystal-ball (worst-case) and crystal-ball (best-case) outcomes. This implies that the proposed strategies capture approximately 60\% of the profit differential between these two extremes. This indicates that, in contrast to the spread trader, a~substantial portion of the exploitable structure in price levels can be captured using the proposed probabilistic forecasting framework. At the same time, the remaining 40\% gap suggests that further improvements may be obtained through enhanced tail modeling and more refined scenario selection mechanisms.

\section{Summary and discussion}
\label{sec:conclusion}

In this paper, we have developed a comprehensive framework for ensemble forecasting of intraday electricity price trajectories and their translation into trading decisions. The empirical results demonstrate that the proposed ensemble forecasting framework yields consistent improvements over benchmark approaches in both statistical and economic terms. In terms of point forecast accuracy, the cSVR-based models achieve lower mean absolute errors than the na\"{i}ve benchmark, confirming the relevance of combining transaction-level information with exogenous variables. At the same time, the probabilistic evaluation reveals a more nuanced picture. While the proposed methods improve the representation of the central part of the predictive distribution, their performance in the tails, as measured by the CRPS, remains slightly inferior to that of the empirical benchmark.

The analysis of forecast accuracy in the case study also highlights systematic differences between scenario construction approaches. Fundamental scenarios, based on dynamics of the forecast errors of system variables, consistently improve trajectory forecast accuracy, indicating their ability to capture structured variations in market conditions. However, they tend to produce more concentrated ensembles, which results in weaker tail performance. In contrast, historical simulation preserves a more comprehensive representation of variability {by capturing idiosyncratic events and non-systemic shocks that drive tail behavior},
thereby yielding superior probabilistic performance. 
These findings point to a trade-off between structural interpretability and distributional richness in scenario-based forecasting.

The proposed Support Vector Sorting procedure plays a key role in improving the efficiency of the ensemble representation. By identifying and retaining only the most influential scenarios, SVS substantially reduces the number of trajectories required to approximate the predictive distribution, while maintaining -- and in some cases improving -- forecast quality. This effect is particularly visible in the tails of historical simulation, suggesting that a relatively small subset of carefully selected scenarios can capture the essential uncertainty structure of the market.

From an economic perspective, the results confirm that model-based probabilistic forecasts translate into tangible gains in trading performance. The na\"{i}ve benchmark ensemble forecast is outperformed in all considered strategy classes in terms of profitability, demonstrating the practical value of incorporating predictive information. 
The comparison between the trading strategies based on different ensemble generation methods reveals a consistent pattern. Strategies based on fundamental scenarios tend to achieve higher profits, albeit at the cost of increased risk, while those based on historical scenarios offer more stable performance with lower downside variability. This trade-off reflects the differing properties of the underlying scenario distributions and suggests that the choice of scenario construction should be aligned with the risk preferences of the market participant.

Furthermore, the best performing trading strategies outperform the considered benchmarks. Dynamic updating through scenario reweighting further enhances performance, leading to higher profits with only a limited impact on downside risk. In the case of seller's strategies, dynamic adjustments can even reduce risk, highlighting the benefits of adaptive decision-making in level-oriented trading contexts. {Finally, the flexibility in parameterizing the weighting function enables the identification of optimal strategy designs tailored to specific agent profiles. Stable, memory-based schemes are most effective in spread trading, while reactive, price-tracking rules prove most beneficial for the seller.}

The methods proposed in this study can be further studied and expanded. First, the dynamic strategy updating mechanism relies solely on deviations between realized price trajectories and scenario paths and does not explicitly account for changes in volatility. Although information on the risk is incorporated at the strategy design stage through the use of the Sortino ratio in parameter selection, the ensemble reweighting itself does not account for risk. Second, the scenario generation approach is deliberately restricted to standard data sources and does not incorporate external meteorological ensemble forecasts. Although this enhances accessibility, it may limit the relevance and richness of the uncertainty representation. 
Finally, the trading strategies are evaluated in a simplified setting that abstracts from transaction costs, market impact, and order book dynamics. 
These limitations point to several directions for future research. In particular, extending the updating mechanism to incorporate volatility-sensitive weighting schemes, enriching scenario generation with external ensemble information, and embedding the framework in more realistic trading environments would further enhance its applicability. 

In general, the results indicate that the combination of kernel-based learning with scenario-driven uncertainty representation provides a flexible and effective approach to probabilistic forecasting in continuous electricity markets. Scenario selection can limit the size of the ensemble without negatively impacting the forecast accuracy. The strong performance of dynamically updated trading strategies underscores the importance of incorporating intra-horizon information and adapting forecasts in real time within continuous trading environments.

\section*{CRediT authorship contribution statement}
\noindent Andrzej Puć: Methodology, Conceptualisation, Formal analysis,  Software, Writing. 
Joanna Janczura: Supervision, Conceptualisation, Resources, Writing.

\section*{Acknowledgments}
\noindent The authors acknowledge the Wrocław Centre for Networking and Supercomputing (WCSS) for providing computational resources that supported this work. The work was partially financed by the Polish National Science Center (NCN) Sonata grant No. 2019/35/D/HS4/00369.

\section*{Data and code availability}
\noindent {The codes can be downloaded from \url{https://github.com/pucandrzej/csvr_scenario_generation}}. The data for all exogenous variables is freely available on the  \cite{entsoe_transparency_platform} platform.  The German intraday continuous transaction data was used under license from the EPEX SPOT. Availability is subject to the EPEX SPOT restrictions.

\section*{Declaration of competing interest}
\noindent The authors declare that they have no known competing financial interests or personal relationships that could have appeared to influence the work reported in this paper.

\bibliographystyle{plainnat}
\bibliography{bibliography}

\appendix
\section{Additional figures}
\begin{figure}[!h]
    \centering
    \includegraphics[width=\textwidth]{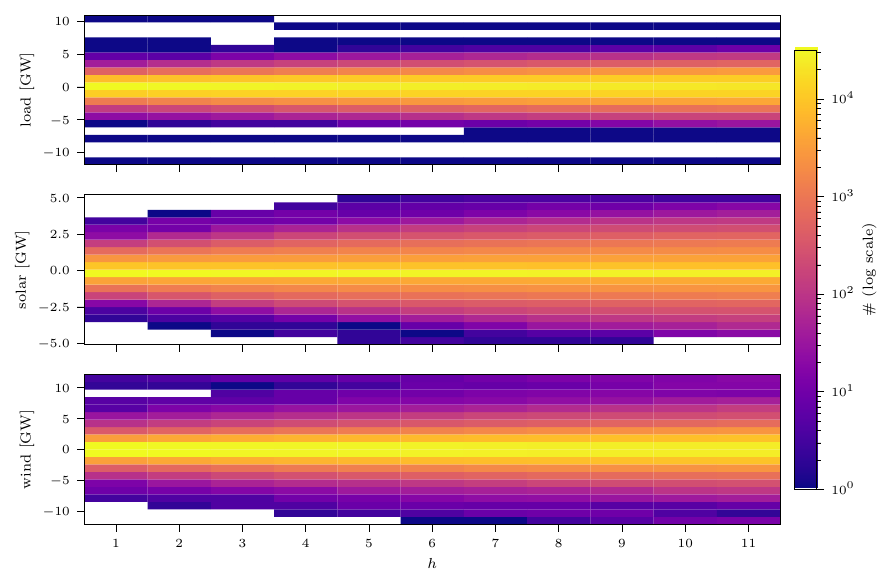}
    \caption{Colormap representing the histograms of fundamental scenarios $\Delta^{(i)}_{\mathrm{fs}}$ for each quarter-hourly step in a path. {Note that, due to different granularity between price and fundamental variables in the case study, the number of path steps differs here}. Bin counts are displayed using a~logarithmic color scale to emphasize variations across several orders of magnitude.}
    \label{fig:scenarios_analysis}
\end{figure}

\begin{figure}[!h]
    \centering
    \includegraphics[width=\textwidth]{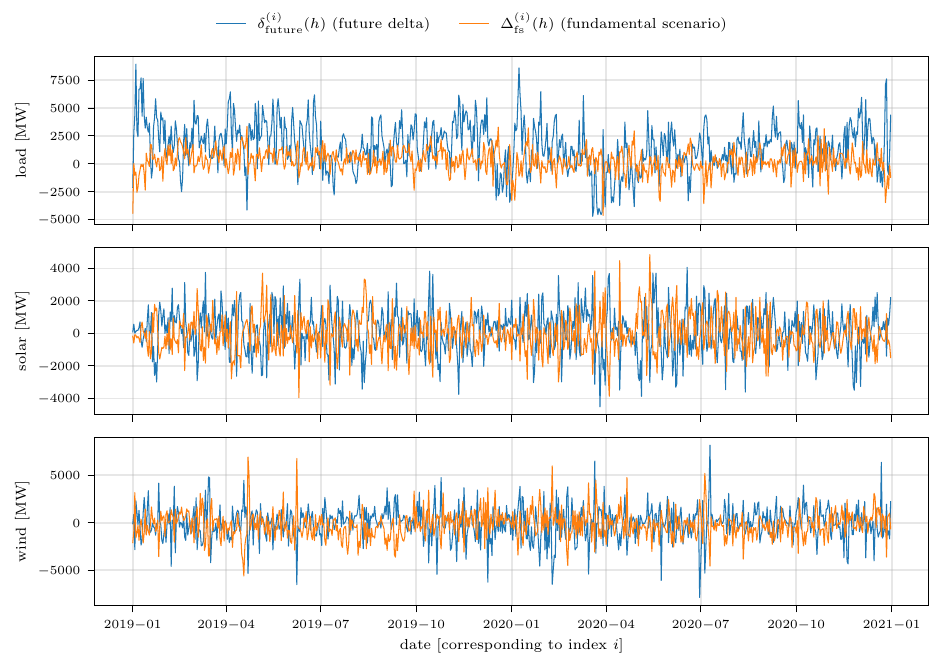}
    \caption{Comparison of future delta and fundamental scenario trajectories for delivery at 14:00 and path step $h=31$.}
    \label{fig:why_double_deltas}
\end{figure}

\clearpage
\section{Additional tables}

\begin{table}[h]
\centering
\resizebox{0.85\textwidth}{!}{%
\begin{tabular}{|ccccc|rrrcrcr|}
\hline
\multicolumn{5}{|c|}{} &
  \multicolumn{7}{c|}{parameters resulting from the grid search} \\ \cline{6-12} 
\multicolumn{5}{|c|}{\multirow{-2}{*}{strategy specification}} &
  \multicolumn{4}{c|}{dynamic kernel} &
  \multicolumn{2}{c|}{dynamic MAE} &
  \multicolumn{1}{c|}{static} \\ \hline
\multicolumn{1}{|c|}{agent} &
  \multicolumn{1}{c|}{strategy} &
  \multicolumn{1}{c|}{\begin{tabular}[c]{@{}c@{}}band\\ type\end{tabular}} &
  \multicolumn{1}{c|}{\begin{tabular}[c]{@{}c@{}}scenario\\ selection\end{tabular}} &
  model &
  \multicolumn{1}{c|}{SCP} &
  \multicolumn{1}{c|}{p} &
  \multicolumn{1}{c|}{lambda} &
  \multicolumn{1}{c|}{threshold} &
  \multicolumn{1}{c|}{SCP} &
  \multicolumn{1}{c|}{threshold} &
  \multicolumn{1}{c|}{SCP} \\ \hline
\multicolumn{1}{|c|}{} &
  \multicolumn{1}{c|}{} &
  \multicolumn{1}{c|}{} &
  \multicolumn{1}{c|}{} &
  na\"{i}ve &
  \multicolumn{1}{r|}{\cellcolor[HTML]{E5E797}0.20} &
  \multicolumn{1}{r|}{\cellcolor[HTML]{A1D289}1.25} &
  \multicolumn{1}{r|}{\cellcolor[HTML]{9ED188}0.35} &
  \multicolumn{1}{c|}{IQR} &
  \multicolumn{1}{r|}{\cellcolor[HTML]{DDE595}0.25} &
  \multicolumn{1}{c|}{5--95 IPR} &
  \cellcolor[HTML]{C3DC90}0.40 \\ \cline{5-12} 
\multicolumn{1}{|c|}{} &
  \multicolumn{1}{c|}{} &
  \multicolumn{1}{c|}{} &
  \multicolumn{1}{c|}{} &
  historical &
  \multicolumn{1}{r|}{\cellcolor[HTML]{DDE595}0.25} &
  \multicolumn{1}{r|}{\cellcolor[HTML]{B6D88D}1.00} &
  \multicolumn{1}{r|}{\cellcolor[HTML]{9ED188}0.35} &
  \multicolumn{1}{c|}{IQR} &
  \multicolumn{1}{r|}{\cellcolor[HTML]{B1D78C}0.50} &
  \multicolumn{1}{c|}{5--95 IPR} &
  \cellcolor[HTML]{DDE595}0.25 \\ \cline{5-12} 
\multicolumn{1}{|c|}{} &
  \multicolumn{1}{c|}{} &
  \multicolumn{1}{c|}{} &
  \multicolumn{1}{c|}{\multirow{-3}{*}{---}} &
  fundamental &
  \multicolumn{1}{r|}{\cellcolor[HTML]{B1D78C}0.50} &
  \multicolumn{1}{r|}{\cellcolor[HTML]{CADF91}0.75} &
  \multicolumn{1}{r|}{\cellcolor[HTML]{9ED188}0.35} &
  \multicolumn{1}{c|}{5--95 IPR} &
  \multicolumn{1}{r|}{\cellcolor[HTML]{B1D78C}0.50} &
  \multicolumn{1}{c|}{5--95 IPR} &
  \cellcolor[HTML]{CBDF91}0.35 \\ \cline{4-12} 
\multicolumn{1}{|c|}{} &
  \multicolumn{1}{c|}{} &
  \multicolumn{1}{c|}{} &
  \multicolumn{1}{c|}{SVS} &
  historical &
  \multicolumn{1}{r|}{\cellcolor[HTML]{B1D78C}0.50} &
  \multicolumn{1}{r|}{\cellcolor[HTML]{DFE596}0.50} &
  \multicolumn{1}{r|}{\cellcolor[HTML]{D8E394}0.20} &
  \multicolumn{1}{c|}{5--95 IPR} &
  \multicolumn{1}{r|}{\cellcolor[HTML]{CBDF91}0.35} &
  \multicolumn{1}{c|}{5--95 IPR} &
  \cellcolor[HTML]{C3DC90}0.40 \\ \cline{4-12} 
\multicolumn{1}{|c|}{} &
  \multicolumn{1}{c|}{} &
  \multicolumn{1}{c|}{\multirow{-5}{*}{\begin{tabular}[c]{@{}c@{}}risk\\ averse\end{tabular}}} &
  \multicolumn{1}{c|}{SVS} &
  fundamental &
  \multicolumn{1}{r|}{\cellcolor[HTML]{7DC781}0.80} &
  \multicolumn{1}{r|}{\cellcolor[HTML]{B6D88D}1.00} &
  \multicolumn{1}{r|}{\cellcolor[HTML]{63BE7B}0.50} &
  \multicolumn{1}{c|}{5--95 IPR} &
  \multicolumn{1}{r|}{\cellcolor[HTML]{75C47F}0.85} &
  \multicolumn{1}{c|}{5--95 IPR} &
  \cellcolor[HTML]{A0D288}0.60 \\ \cline{3-12} 
\multicolumn{1}{|c|}{} &
  \multicolumn{1}{c|}{} &
  \multicolumn{1}{c|}{} &
  \multicolumn{1}{c|}{} &
   na\"{i}ve &
  \multicolumn{1}{r|}{\cellcolor[HTML]{CBDF91}0.35} &
  \multicolumn{1}{r|}{\cellcolor[HTML]{B6D88D}1.00} &
  \multicolumn{1}{r|}{\cellcolor[HTML]{8ACB84}0.40} &
  \multicolumn{1}{c|}{IQR} &
  \multicolumn{1}{r|}{\cellcolor[HTML]{EEEA99}0.15} &
  \multicolumn{1}{c|}{IQR} &
  \cellcolor[HTML]{8FCC85}0.70 \\ \cline{5-12} 
\multicolumn{1}{|c|}{} &
  \multicolumn{1}{c|}{} &
  \multicolumn{1}{c|}{} &
  \multicolumn{1}{c|}{} &
  historical &
  \multicolumn{1}{r|}{\cellcolor[HTML]{EEEA99}0.15} &
  \multicolumn{1}{r|}{\cellcolor[HTML]{CADF91}0.75} &
  \multicolumn{1}{r|}{\cellcolor[HTML]{63BE7B}0.50} &
  \multicolumn{1}{c|}{IQR} &
  \multicolumn{1}{r|}{\cellcolor[HTML]{F7ED9B}0.10} &
  \multicolumn{1}{c|}{5--95 IPR} &
  \cellcolor[HTML]{E5E797}0.20 \\ \cline{5-12} 
\multicolumn{1}{|c|}{} &
  \multicolumn{1}{c|}{} &
  \multicolumn{1}{c|}{} &
  \multicolumn{1}{c|}{\multirow{-3}{*}{---}} &
  fundamental &
  \multicolumn{1}{r|}{\cellcolor[HTML]{CBDF91}0.35} &
  \multicolumn{1}{r|}{\cellcolor[HTML]{DFE596}0.50} &
  \multicolumn{1}{r|}{\cellcolor[HTML]{63BE7B}0.50} &
  \multicolumn{1}{c|}{5--95 IPR} &
  \multicolumn{1}{r|}{\cellcolor[HTML]{EEEA99}0.15} &
  \multicolumn{1}{c|}{5--95 IPR} &
  \cellcolor[HTML]{F7ED9B}0.10 \\ \cline{4-12} 
\multicolumn{1}{|c|}{} &
  \multicolumn{1}{c|}{} &
  \multicolumn{1}{c|}{} &
  \multicolumn{1}{c|}{SVS} &
  historical &
  \multicolumn{1}{r|}{\cellcolor[HTML]{DDE595}0.25} &
  \multicolumn{1}{r|}{\cellcolor[HTML]{78C580}1.75} &
  \multicolumn{1}{r|}{\cellcolor[HTML]{9ED188}0.35} &
  \multicolumn{1}{c|}{IQR} &
  \multicolumn{1}{r|}{\cellcolor[HTML]{DDE595}0.25} &
  \multicolumn{1}{c|}{5--95 IPR} &
  \cellcolor[HTML]{DDE595}0.25 \\ \cline{4-12} 
\multicolumn{1}{|c|}{} &
  \multicolumn{1}{c|}{\multirow{-10}{*}{bands}} &
  \multicolumn{1}{c|}{\multirow{-5}{*}{\begin{tabular}[c]{@{}c@{}}risk\\ seeking\end{tabular}}} &
  \multicolumn{1}{c|}{SVS} &
  fundamental &
  \multicolumn{1}{r|}{\cellcolor[HTML]{DDE595}0.25} &
  \multicolumn{1}{r|}{\cellcolor[HTML]{DFE596}0.50} &
  \multicolumn{1}{r|}{\cellcolor[HTML]{9ED188}0.35} &
  \multicolumn{1}{c|}{5--95 IPR} &
  \multicolumn{1}{r|}{\cellcolor[HTML]{DDE595}0.25} &
  \multicolumn{1}{c|}{5--95 IPR} &
  \cellcolor[HTML]{DDE595}0.25 \\ \cline{2-12} 
\multicolumn{1}{|c|}{} &
  \multicolumn{1}{c|}{} &
  \multicolumn{1}{c|}{} &
  \multicolumn{1}{c|}{} &
  na\"{i}ve &
  \multicolumn{1}{r|}{---} &
  \multicolumn{1}{r|}{\cellcolor[HTML]{FFEF9C}0.10} &
  \multicolumn{1}{r|}{\cellcolor[HTML]{77C580}0.45} &
  \multicolumn{1}{c|}{5--95 IPR} &
  \multicolumn{1}{r|}{---} &
  \multicolumn{1}{c|}{5--95 IPR} &
  --- \\ \cline{5-12} 
\multicolumn{1}{|c|}{} &
  \multicolumn{1}{c|}{} &
  \multicolumn{1}{c|}{} &
  \multicolumn{1}{c|}{} &
  historical &
  \multicolumn{1}{r|}{---} &
  \multicolumn{1}{r|}{\cellcolor[HTML]{DFE596}0.50} &
  \multicolumn{1}{r|}{\cellcolor[HTML]{63BE7B}0.50} &
  \multicolumn{1}{c|}{5--95 IPR} &
  \multicolumn{1}{r|}{---} &
  \multicolumn{1}{c|}{5--95 IPR} &
  --- \\ \cline{5-12} 
\multicolumn{1}{|c|}{} &
  \multicolumn{1}{c|}{} &
  \multicolumn{1}{c|}{} &
  \multicolumn{1}{c|}{\multirow{-3}{*}{---}} &
  fundamental &
  \multicolumn{1}{r|}{---} &
  \multicolumn{1}{r|}{\cellcolor[HTML]{DFE596}0.50} &
  \multicolumn{1}{r|}{\cellcolor[HTML]{63BE7B}0.50} &
  \multicolumn{1}{c|}{5--95 IPR} &
  \multicolumn{1}{r|}{---} &
  \multicolumn{1}{c|}{5--95 IPR} &
  --- \\ \cline{4-12} 
\multicolumn{1}{|c|}{} &
  \multicolumn{1}{c|}{} &
  \multicolumn{1}{c|}{} &
  \multicolumn{1}{c|}{SVS} &
  historical &
  \multicolumn{1}{r|}{---} &
  \multicolumn{1}{r|}{\cellcolor[HTML]{F3EC9A}0.25} &
  \multicolumn{1}{r|}{\cellcolor[HTML]{77C580}0.45} &
  \multicolumn{1}{c|}{$3\sigma$} &
  \multicolumn{1}{r|}{---} &
  \multicolumn{1}{c|}{5--95 IPR} &
  --- \\ \cline{4-12} 
\multicolumn{1}{|c|}{\multirow{-15}{*}{\begin{tabular}[c]{@{}c@{}}spread\\ trader\end{tabular}}} &
  \multicolumn{1}{c|}{\multirow{-5}{*}{median}} &
  \multicolumn{1}{c|}{\multirow{-5}{*}{---}} &
  \multicolumn{1}{c|}{SVS} &
  fundamental &
  \multicolumn{1}{r|}{---} &
  \multicolumn{1}{r|}{\cellcolor[HTML]{B6D88D}1.00} &
  \multicolumn{1}{r|}{\cellcolor[HTML]{D8E394}0.20} &
  \multicolumn{1}{c|}{5--95 IPR} &
  \multicolumn{1}{r|}{---} &
  \multicolumn{1}{c|}{5--95 IPR} &
  --- \\ \hline
\multicolumn{1}{|c|}{} &
  \multicolumn{1}{c|}{} &
  \multicolumn{1}{c|}{} &
  \multicolumn{1}{c|}{} &
  na\"{i}ve &
  \multicolumn{1}{r|}{\cellcolor[HTML]{6CC17D}0.90} &
  \multicolumn{1}{r|}{\cellcolor[HTML]{78C580}1.75} &
  \multicolumn{1}{r|}{\cellcolor[HTML]{63BE7B}0.50} &
  \multicolumn{1}{c|}{IQR} &
  \multicolumn{1}{r|}{\cellcolor[HTML]{63BE7B}0.95} &
  \multicolumn{1}{c|}{IQR} &
  \cellcolor[HTML]{6CC17D}0.90 \\ \cline{5-12} 
\multicolumn{1}{|c|}{} &
  \multicolumn{1}{c|}{} &
  \multicolumn{1}{c|}{} &
  \multicolumn{1}{c|}{} &
  historical &
  \multicolumn{1}{r|}{\cellcolor[HTML]{63BE7B}0.95} &
  \multicolumn{1}{r|}{\cellcolor[HTML]{CADF91}0.75} &
  \multicolumn{1}{r|}{\cellcolor[HTML]{9ED188}0.35} &
  \multicolumn{1}{c|}{IQR} &
  \multicolumn{1}{r|}{\cellcolor[HTML]{63BE7B}0.95} &
  \multicolumn{1}{c|}{IQR} &
  \cellcolor[HTML]{63BE7B}0.95 \\ \cline{5-12} 
\multicolumn{1}{|c|}{} &
  \multicolumn{1}{c|}{} &
  \multicolumn{1}{c|}{} &
  \multicolumn{1}{c|}{\multirow{-3}{*}{---}} &
  fundamental &
  \multicolumn{1}{r|}{\cellcolor[HTML]{63BE7B}0.95} &
  \multicolumn{1}{r|}{\cellcolor[HTML]{A1D289}1.25} &
  \multicolumn{1}{r|}{\cellcolor[HTML]{FFEF9C}0.10} &
  \multicolumn{1}{c|}{IQR} &
  \multicolumn{1}{r|}{\cellcolor[HTML]{63BE7B}0.95} &
  \multicolumn{1}{c|}{IQR} &
  \cellcolor[HTML]{63BE7B}0.95 \\ \cline{4-12} 
\multicolumn{1}{|c|}{} &
  \multicolumn{1}{c|}{} &
  \multicolumn{1}{c|}{} &
  \multicolumn{1}{c|}{SVS} &
  historical &
  \multicolumn{1}{r|}{\cellcolor[HTML]{63BE7B}0.95} &
  \multicolumn{1}{r|}{\cellcolor[HTML]{DFE596}0.50} &
  \multicolumn{1}{r|}{\cellcolor[HTML]{FFEF9C}0.10} &
  \multicolumn{1}{c|}{IQR} &
  \multicolumn{1}{r|}{\cellcolor[HTML]{63BE7B}0.95} &
  \multicolumn{1}{c|}{IQR} &
  \cellcolor[HTML]{63BE7B}0.95 \\ \cline{4-12} 
\multicolumn{1}{|c|}{} &
  \multicolumn{1}{c|}{} &
  \multicolumn{1}{c|}{\multirow{-5}{*}{\begin{tabular}[c]{@{}c@{}}risk\\ averse\end{tabular}}} &
  \multicolumn{1}{c|}{SVS} &
  fundamental &
  \multicolumn{1}{r|}{\cellcolor[HTML]{63BE7B}0.95} &
  \multicolumn{1}{r|}{\cellcolor[HTML]{A1D289}1.25} &
  \multicolumn{1}{r|}{\cellcolor[HTML]{63BE7B}0.50} &
  \multicolumn{1}{c|}{IQR} &
  \multicolumn{1}{r|}{\cellcolor[HTML]{63BE7B}0.95} &
  \multicolumn{1}{c|}{IQR} &
  \cellcolor[HTML]{63BE7B}0.95 \\ \cline{3-12} 
\multicolumn{1}{|c|}{} &
  \multicolumn{1}{c|}{} &
  \multicolumn{1}{c|}{} &
  \multicolumn{1}{c|}{} &
  na\"{i}ve &
  \multicolumn{1}{r|}{\cellcolor[HTML]{FFEF9C}0.05} &
  \multicolumn{1}{r|}{\cellcolor[HTML]{78C580}1.75} &
  \multicolumn{1}{r|}{\cellcolor[HTML]{9ED188}0.35} &
  \multicolumn{1}{c|}{5--95 IPR} &
  \multicolumn{1}{r|}{\cellcolor[HTML]{FFEF9C}0.05} &
  \multicolumn{1}{c|}{IQR} &
  \cellcolor[HTML]{FFEF9C}0.05 \\ \cline{5-12} 
\multicolumn{1}{|c|}{} &
  \multicolumn{1}{c|}{} &
  \multicolumn{1}{c|}{} &
  \multicolumn{1}{c|}{} &
  historical &
  \multicolumn{1}{r|}{\cellcolor[HTML]{FFEF9C}0.05} &
  \multicolumn{1}{r|}{\cellcolor[HTML]{DFE596}0.50} &
  \multicolumn{1}{r|}{\cellcolor[HTML]{9ED188}0.35} &
  \multicolumn{1}{c|}{IQR} &
  \multicolumn{1}{r|}{\cellcolor[HTML]{FFEF9C}0.05} &
  \multicolumn{1}{c|}{IQR} &
  \cellcolor[HTML]{FFEF9C}0.05 \\ \cline{5-12} 
\multicolumn{1}{|c|}{} &
  \multicolumn{1}{c|}{} &
  \multicolumn{1}{c|}{} &
  \multicolumn{1}{c|}{\multirow{-3}{*}{---}} &
  fundamental &
  \multicolumn{1}{r|}{\cellcolor[HTML]{FFEF9C}0.05} &
  \multicolumn{1}{r|}{\cellcolor[HTML]{A1D289}1.25} &
  \multicolumn{1}{r|}{\cellcolor[HTML]{9ED188}0.35} &
  \multicolumn{1}{c|}{IQR} &
  \multicolumn{1}{r|}{\cellcolor[HTML]{FFEF9C}0.05} &
  \multicolumn{1}{c|}{IQR} &
  \cellcolor[HTML]{FFEF9C}0.05 \\ \cline{4-12} 
\multicolumn{1}{|c|}{} &
  \multicolumn{1}{c|}{} &
  \multicolumn{1}{c|}{} &
  \multicolumn{1}{c|}{SVS} &
  historical &
  \multicolumn{1}{r|}{\cellcolor[HTML]{FFEF9C}0.05} &
  \multicolumn{1}{r|}{\cellcolor[HTML]{DFE596}0.50} &
  \multicolumn{1}{r|}{\cellcolor[HTML]{FFEF9C}0.10} &
  \multicolumn{1}{c|}{IQR} &
  \multicolumn{1}{r|}{\cellcolor[HTML]{FFEF9C}0.05} &
  \multicolumn{1}{c|}{IQR} &
  \cellcolor[HTML]{FFEF9C}0.05 \\ \cline{4-12} 
\multicolumn{1}{|c|}{} &
  \multicolumn{1}{c|}{\multirow{-10}{*}{bands}} &
  \multicolumn{1}{c|}{\multirow{-5}{*}{\begin{tabular}[c]{@{}c@{}}risk\\ seeking\end{tabular}}} &
  \multicolumn{1}{c|}{SVS} &
  fundamental &
  \multicolumn{1}{r|}{\cellcolor[HTML]{FFEF9C}0.05} &
  \multicolumn{1}{r|}{\cellcolor[HTML]{B6D88D}1.00} &
  \multicolumn{1}{r|}{\cellcolor[HTML]{FFEF9C}0.10} &
  \multicolumn{1}{c|}{IQR} &
  \multicolumn{1}{r|}{\cellcolor[HTML]{FFEF9C}0.05} &
  \multicolumn{1}{c|}{IQR} &
  \cellcolor[HTML]{F7ED9B}0.10 \\ \cline{2-12} 
\multicolumn{1}{|c|}{} &
  \multicolumn{1}{c|}{} &
  \multicolumn{1}{c|}{} &
  \multicolumn{1}{c|}{} &
  na\"{i}ve &
  \multicolumn{1}{r|}{---} &
  \multicolumn{1}{r|}{\cellcolor[HTML]{B6D88D}1.00} &
  \multicolumn{1}{r|}{\cellcolor[HTML]{8ACB84}0.40} &
  \multicolumn{1}{c|}{IQR} &
  \multicolumn{1}{r|}{---} &
  \multicolumn{1}{c|}{$3\sigma$} &
  --- \\ \cline{5-12} 
\multicolumn{1}{|c|}{} &
  \multicolumn{1}{c|}{} &
  \multicolumn{1}{c|}{} &
  \multicolumn{1}{c|}{} &
  historical &
  \multicolumn{1}{r|}{---} &
  \multicolumn{1}{r|}{\cellcolor[HTML]{63BE7B}2.00} &
  \multicolumn{1}{r|}{\cellcolor[HTML]{63BE7B}0.50} &
  \multicolumn{1}{c|}{IQR} &
  \multicolumn{1}{r|}{---} &
  \multicolumn{1}{c|}{$3\sigma$} &
  --- \\ \cline{5-12} 
\multicolumn{1}{|c|}{} &
  \multicolumn{1}{c|}{} &
  \multicolumn{1}{c|}{} &
  \multicolumn{1}{c|}{\multirow{-3}{*}{---}} &
  fundamental &
  \multicolumn{1}{r|}{---} &
  \multicolumn{1}{r|}{\cellcolor[HTML]{A1D289}1.25} &
  \multicolumn{1}{r|}{\cellcolor[HTML]{FFEF9C}0.10} &
  \multicolumn{1}{c|}{5--95 IPR} &
  \multicolumn{1}{r|}{---} &
  \multicolumn{1}{c|}{5--95 IPR} &
  --- \\ \cline{4-12} 
\multicolumn{1}{|c|}{} &
  \multicolumn{1}{c|}{} &
  \multicolumn{1}{c|}{} &
  \multicolumn{1}{c|}{SVS} &
  historical &
  \multicolumn{1}{r|}{---} &
  \multicolumn{1}{r|}{\cellcolor[HTML]{78C580}1.75} &
  \multicolumn{1}{r|}{\cellcolor[HTML]{B2D78C}0.30} &
  \multicolumn{1}{c|}{IQR} &
  \multicolumn{1}{r|}{---} &
  \multicolumn{1}{c|}{5--95 IPR} &
  --- \\ \cline{4-12} 
\multicolumn{1}{|c|}{\multirow{-15}{*}{seller}} &
  \multicolumn{1}{c|}{\multirow{-5}{*}{median}} &
  \multicolumn{1}{c|}{\multirow{-5}{*}{---}} &
  \multicolumn{1}{c|}{SVS} &
  fundamental &
  \multicolumn{1}{r|}{---} &
  \multicolumn{1}{r|}{\cellcolor[HTML]{A1D289}1.25} &
  \multicolumn{1}{r|}{\cellcolor[HTML]{FFEF9C}0.10} &
  \multicolumn{1}{c|}{5--95 IPR} &
  \multicolumn{1}{r|}{---} &
  \multicolumn{1}{c|}{5--95 IPR} &
  --- \\ \hline
\end{tabular}
}
\caption{{Hyperparameters of trading strategies obtained from the grid search applied separately for each strategy specification.} Color scales are separate for each parameter.}
\label{tab:strrategies_grid_search_results}
\end{table}

\end{document}